\documentclass[iop,apj]{emulateapj}

\usepackage{natbib}
\usepackage{afterpage}
\usepackage{times}
\usepackage{amsmath,amssymb,amstext}
\usepackage{rotating}

\usepackage[colorlinks=blue,linkcolor=blue,urlcolor=blue,citecolor=blue]{hyperref} 

\usepackage[all]{hypcap}
\usepackage{tabularx}
\usepackage{float}
\usepackage{natbib}
\usepackage[usenames, dvipsnames]{color}

\newcommand{\kms}[1]{$#1$~km~s$^{-1}$}

\begin{document}

\title{Observation of a Large-scale Quasi-circular Secondary Ribbon associated with Successive Flares and a Halo CME}

\author{Eun-Kyung Lim\altaffilmark{1}, Vasyl Yurchyshyn\altaffilmark{1,2}, Pankaj Kumar\altaffilmark{1,3}, Kyuhyoun Cho\altaffilmark{4}, Chaowei Jiang\altaffilmark{5}, Sujin Kim\altaffilmark{1}, Heesu Yang\altaffilmark{1}, Jongchul Chae\altaffilmark{4}, Kyung-Suk Cho\altaffilmark{1,6}, and Jeongwoo Lee\altaffilmark{4}}
\affil{$^1$Korea Astronomy and Space Science Institute 776, Daedeokdae-ro, Yuseong-gu, Daejeon, Republic of Korea 305-348}
\affil{$^2$Big Bear Solar Observatory, New Jersey Institute of Technology, 40386 North Shore Lane, Big Bear City, CA 92314-9672, USA}
\affil{$^3$Heliophysics Science Division, NASA Goddard Space Flight Center, Greenbelt, MD, 20771, USA}
\affil{$^4$Astronomy Program, Department of Physics and Astronomy, Seoul National University, Seoul 151-747, Korea}
\affil{$^5$Institute of Space Science and Applied Technology, Harbin Institute of Technology, Shenzhen 518055, China}
\affil{$^6$University of Science and Technology, Daejeon 305-348, Korea}
\email{eklim@kasi.re.kr}


\begin{abstract}
{Solar flare ribbons provide an important clue to the magnetic reconnection process and associated magnetic field topology in the solar corona. We detected a large-scale secondary flare ribbon of a circular shape that developed in association with two successive M-class flares and one CME. The ribbon revealed interesting properties such as 1) a quasi-circular shape and enclosing the central active region; 2) the size as large as 500\arcsec\, by 650\arcsec\,, 3) successive brightenings in the clockwise direction at a speed of \kms{160} starting from the nearest position to the flaring sunspots, 4) radial contraction and expansion in the northern and the southern part, respectively at speeds of $\leq$ \kms{10}. Using multi-wavelength data from \textit{SDO}, \textit{RHESSI}, XRT, and Nobeyama, along with magnetic field extrapolations, we found that: 1) the secondary ribbon location is consistent with the field line footpoints of a fan-shaped magnetic structure that connects the flaring region and the ambient decaying field; 2) the second M2.6 flare occurred when the expanding coronal loops driven by the first M2.0 flare encountered the background decayed field. 3) Immediately after the second flare, the secondary ribbon developed along with dimming regions. Based on our findings, we suggest that interaction between the expanding sigmoid field and the overlying fan-shaped field triggered the secondary reconnection that resulted in the field opening and formation of the quasi-circular secondary ribbon. We thus conclude that interaction between the active region and the ambient large-scale fields should be taken into account to fully understand the entire eruption process.}
\end{abstract}


\section{INTRODUCTION}
Flare ribbons are one of the representative signatures of solar flares and are now recognized as chromospheric footpoints of coronal loops that experienced magnetic reconnection \citep{Stu68}. Ribbons are detected as elongated bright structures in almost all wavelengths of the solar spectrum, most noticeably in the H$\alpha$ and ultraviolet passbands. Their emission is explained by the precipitation of accelerated particles from the reconnection site and the thermal conduction from flaring coronal loops \citep{Fle01}. Based on morphological and dynamical studies of flare ribbons, important information on both magnetic reconnection and the involved topology of the magnetic field can be obtained.

The typical appearance of flare ribbons is a pair of two parallel threads, so called two-ribbon structure, which are frequently associated with eruptive flares that are mostly long durational events. The two-ribbon  structure shows extension of brightening along the ribbon axis as well as apparent separation from each other in time. These phenomena are well explained in terms of magnetic reconnection between opposite directed open fields in an X-shaped null-point configuration used in the standard flare model \citep[CSHKP model;][]{Car64, Stu66, Hir74, Kop76}. As reconnection between the merging open fields proceeds, the height of the reconnection site increases and thus the field lines in greater distance go through the reconnection resulting in apparent separation of flare ribbons farther away from each other \citep{Kop76}.

Another type of a flare ribbon that is recently actively studied is a circular flare ribbon \citep{Mas09}. This type of a ribbon shaped as a closed curve is explained in terms of 3-dimensional (3D) null-point reconnection with the magnetic field topology of a fan surface and spines \citep{Lau90, Pon07}. This 3D reconnection involves fan and spine-reconnection, as well as a hybrid, so called spine-fan reconnection \citep{Pri09}. In addition, depending on the driver, torsional reconnection is also suggested \citep{Pon07}. Such phenomena involve a parasitic polarity surrounded by opposed polarity fields resulting in the circular shape of the polarity inversion line. The main observational signatures of circular flare ribbons are brightenings at the foopoints of the fan-shaped field lines as obtained from field extrapolations based on either potential or non-linear force-free field (NLFFF) models, and the sequential brightening propagating along the ribbon. \citet{Mas09} reproduced this pattern of the sequential brightenings in their magnetohydrodynamics (MHD) simulations with an asymmetric fan configuration in terms of field slippage toward the null-point prior to the spine-reconnection. Due to the slower (sub-Alfv$\acute{e}$nic) slippage speed of the field lines that are farther away from the null-point compared to the faster (super-Alfv$\acute{e}$nic) slippage of the closer field lines, the reconnection of the farther lines takes place later thus resulting in delayed brightenings at footpoints of the remote field lines.

The concept of quasi-separatrix layers (QSLs) is frequently used to interpret flare events with a complex ribbon structure. The QSL is defined as a narrow volume across which the connectivity of the magnetic field drastically changes and the electric current is thus enhanced although the field is still continuous \citep{Pri95, Dem96a, Dem96b}. The concept of QSLs was first proposed by \citet{Pri95} in order to explain magnetic reconnection in a magnetic topology without true separatrices. According to \citet{Pri92}'s suggestion, in a 3D magnetic reconnection without separatrices, magnetic field lines should slip along each other within a certain layer that was further developed into a QSL. Complex deformation of QSLs and thus complex current layers are expected to easily occur in a case, for example, when one quasi-connectivity domain is expanded and the other domain is shrunk by the boundary motions \citep{Aul06}.

Recently, \citet{Zha14} studied 10 X-class flares and found that 11 flares revealed multiple flare ribbons indicating a complex reconnection process rather than the standard flare model reconnection. By classifying those multiple flare ribbons into two types, these authors firstly introduced a term secondary flare ribbons, that are characterized by weaker intensity than that of the normal flare ribbons, absence of associated post-flare loops, and a short lifetime. Although the morphology of each secondary flare ribbon was not subject of their study, the sample images they provided showed that observed secondary flare ribbons are either two-ribbon or single arc shaped. A large-scale secondary magnetic reconnection near the QSL driven by the initial reconnection of the main flare and/or by the flare blast waves was suggested as their formation mechanism.

In this study we present data on formation of a large-scale flare ribbon of a quasi-circular shape. Based on the observed characteristics of the ribbon that are similar to those reported by \citet{Zha14} except for the lifetime, we identified this ribbon as a ``quasi-circular secondary ribbon (QCSR)". This ribbon appeared in association with two subsequent M-class flares and a halo coronal mass ejection (CME). Related observed events such as loop expansion, plasma ejection, coronal dimming, etc. were well detected by the \textit{Solar Dynamics Observatory} \citep[SDO;][]{Pes12} instruments in addition to the ribbon development. We analyzed dynamical properties of these observed phenomena in order to understand the entire eruption process and the formation mechanism of the secondary flare ribbon.


\section{OBSERVATIONS AND DATA ANALYSIS}\label{obs and data}
\begin{figure}[tb]
\begin{center}
    \includegraphics[width=0.48\textwidth]{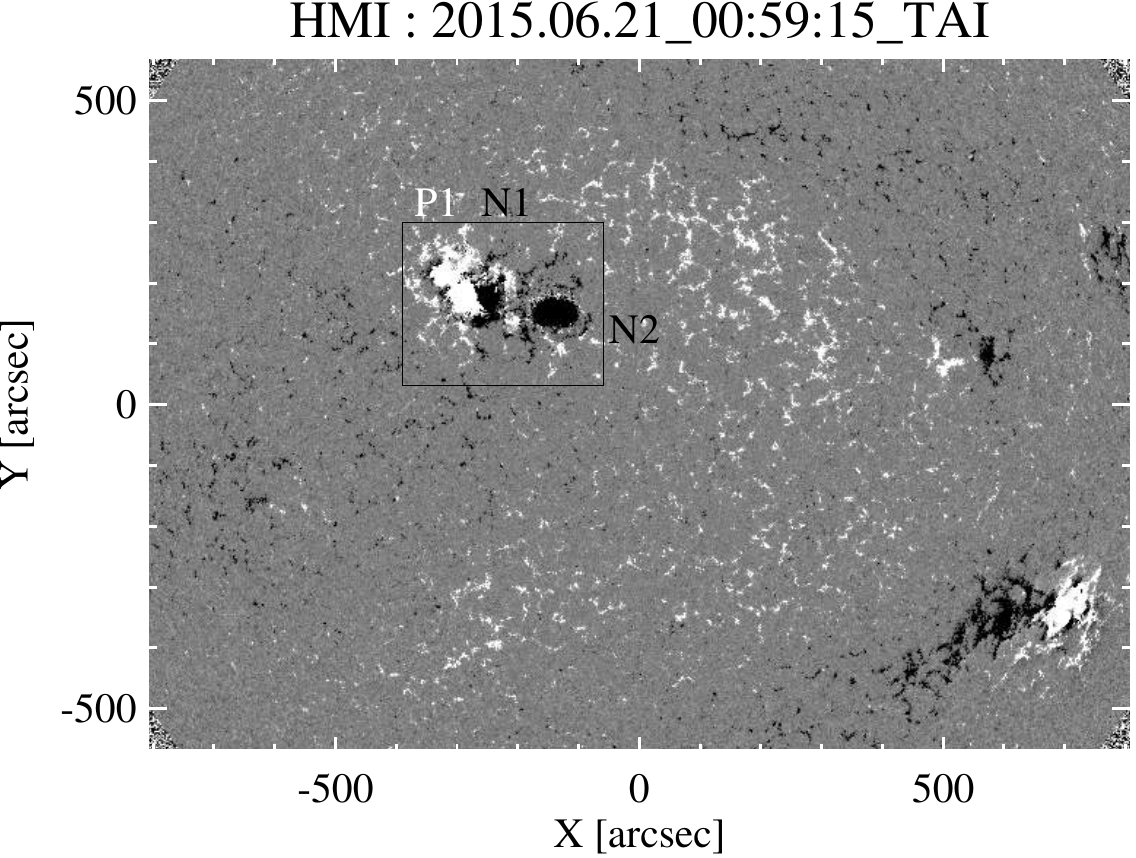}
    \includegraphics[width=0.48\textwidth]{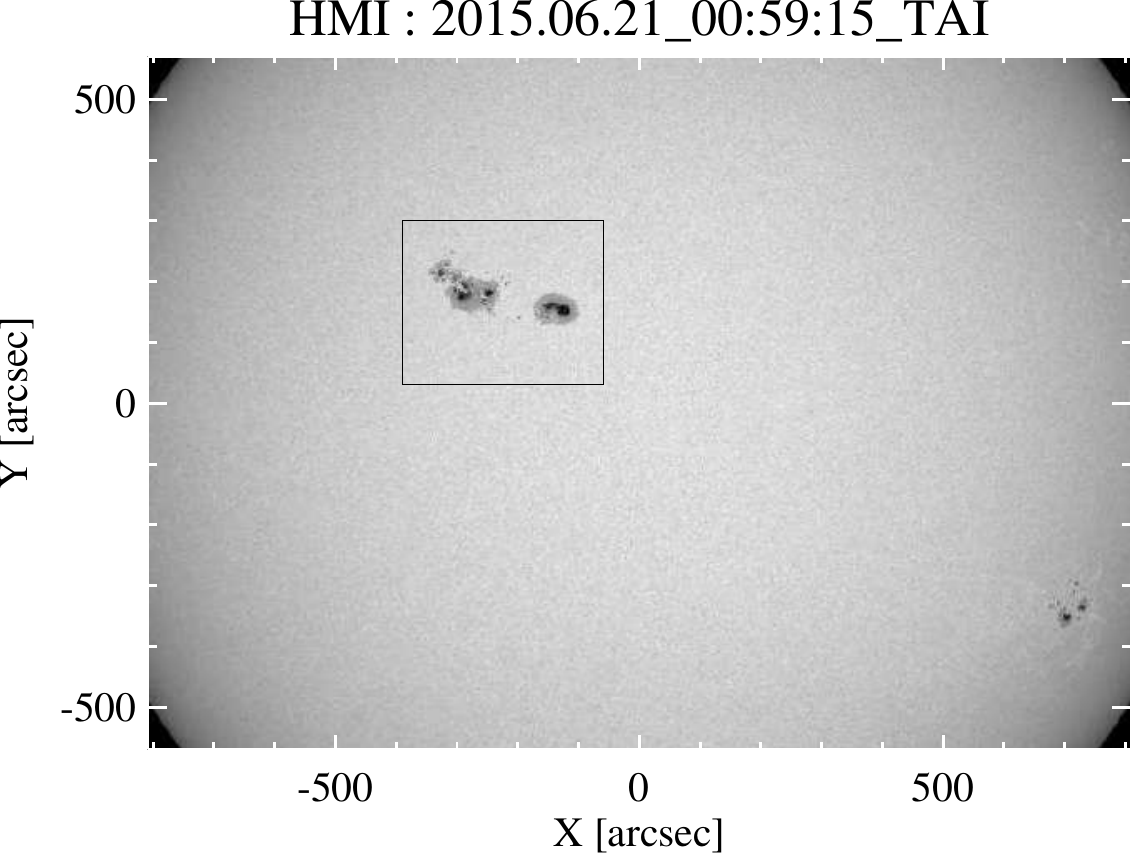}\\
    \includegraphics[width=0.48\textwidth]{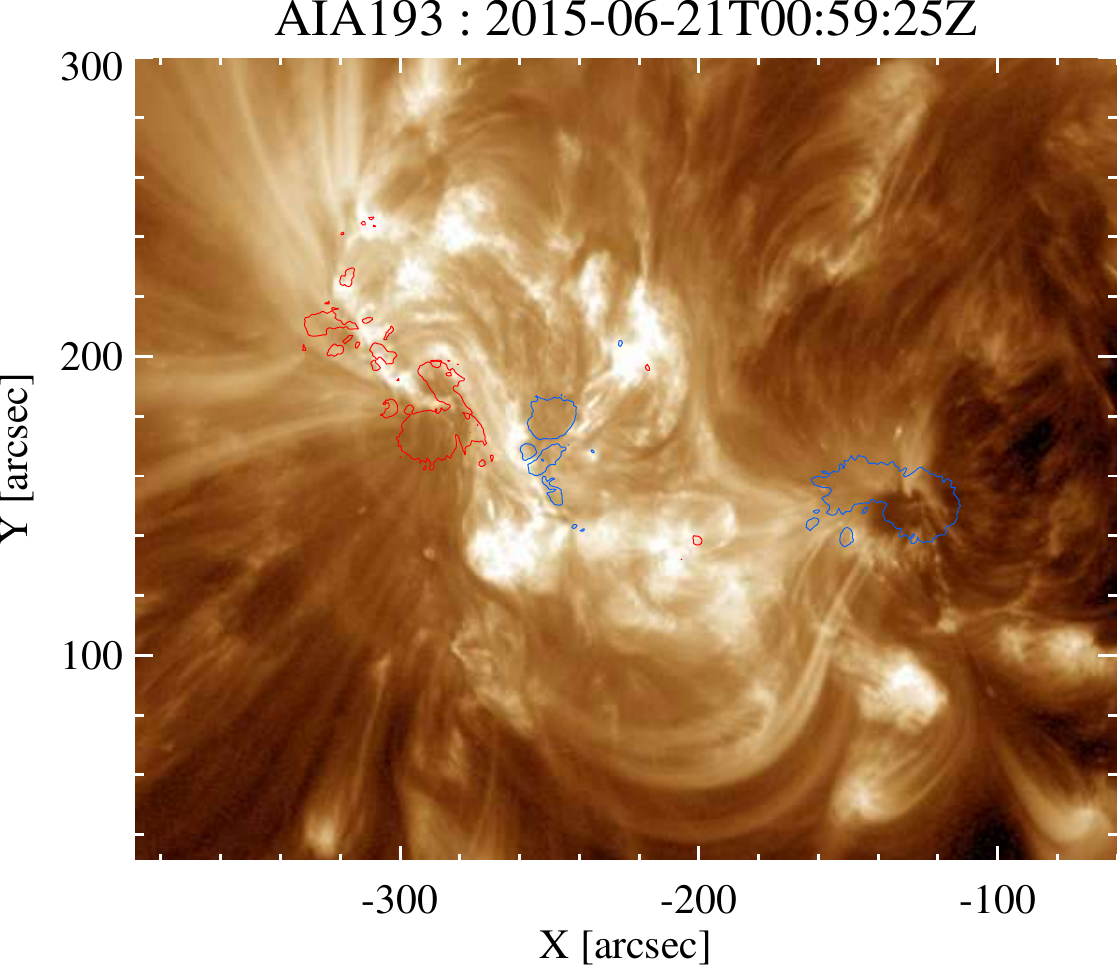}
    \includegraphics[width=0.48\textwidth]{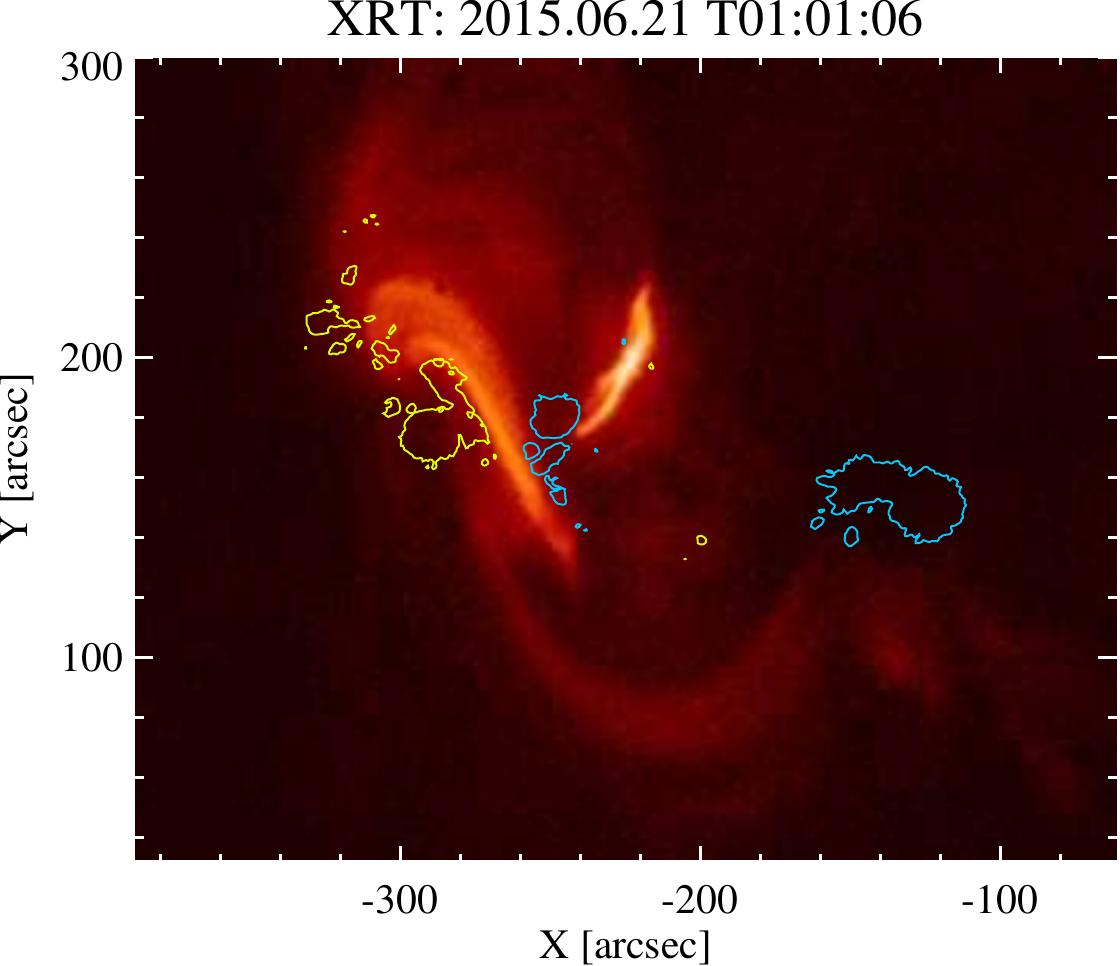}\\
    \caption{Top: SDO/HMI magnetogram (left) and intensity map (right) of active region NOAA 12371 taken at 00:59:15~UT on 2015 June 21. Bottom: AIA~193\AA\ (left) image and the Hinode/XRT (right) coronal images of the region outlined by the black box in upper panels. The magnetic field contours are plotted at level of $\pm1000$~G with red (yellow) and blue (cyan) colors, respectively.}\label{hmi_full}
\end{center}
\end{figure}

NOAA active region (AR) 12371 located at (-220\arcsec, 158\arcsec) on 2015 June 21 was the only flare productive AR near the disk center during our time of interest (Figure~\ref{hmi_full}). During the disk passage from June 16 to 27, this AR produced six M-class and over 30 C-class flares among which the M7.9 flare was the strongest. On June 21, the AR produced two M-class flares within one-hour time interval. The first M2.0 flare, referred to as Flare I hereafter, started at 01:02:00~UT and the M2.6 flare (Flare II), started at 02:06:00~UT. These flares are focus of our attention because of their dynamic evolution as successive events and their intriguing observational biproducts, such as a halo CME and a strikingly large, QCSR partly enclosing the flaring AR.

We utilized \textit{SDO} Atmospheric Imaging Assembly \citep[AIA;][]{Lem12} EUV data to study the overall process of AR eruption including the two flares and one halo CME. The \emph{SDO}/AIA uninterruptedly provides full-disk images of the transition region and the corona in multiple spectral channels with high spatial resolution of 1.5 arcsec and temporal cadence of 12 second. Among seven extreme ultraviolet (EUV) channels, we used the 193~\AA\ (1.58~MK, 20~MK), 131~\AA\ (0.40~MK, 10.00~MK, 16~MK) and 171~\AA\ (0.63~MK) data to analyze temporal changes in coronal structures driven by the flares, and the 304~\AA\ (0.05~MK) data for QCSR studies. To identify and trace the temporal evolution of the normal two-ribbon structure, we used the AIA~1600~\AA\ (0.1~MK) data. The X-ray data captured by the \emph{Geostationary Orbiting Environmental Satellites (GOES)} and \emph{Reuven Ramaty High Energy Solar Spectroscopic Imager} \citep[\emph{REHSSI};][]{Lin02} were also used to examine the temporal change of energy release during flares.

We used the solar surface magnetic field data from the \emph{SDO}/\emph{Helioseismic and Magnetic Imager} \citep[HMI;][]{Sch12}, in particular, the Space weather HMI Active Region Patches (SHARP) vector magnetogram data series \citep{Hoe14, Bob14} for NLFFF extrapolations. We carried out two types of extrapolations, one is based on a potential field approach and the other on a NLFFF method. The region of interest includes a single AR surrounded by wide areas of background decayed fields, and hence the region of interest is too wide to be reconstructed by most of the existing NLFFF extrapolation techniques. The potential field approximation not only saves the computational time but also gives reasonable physical insights into the structure of the large-scale coronal field. For this purpose we used the potential-field source-surface package \citep[PFSS; e.g.,][]{Sch69} available through the SolarSoftware (SSW). Pre-computed PFSS models with a 6-hour sampling are available via download through the SSW, and these models are calculated using the scheme described on \citet{Sch03}. In this study, we utilized a 01:00:00~UT PFSS model on June 21, a few minutes before the flare I. We also used the NLFFF model developed by \citet{Jia12} to reveal the nonpotential nature of the flaring region. The model is based on an MHD relaxation method and the space-time conservation-element/solution-element method, and it has been improved by \citet{Jia13} to be able to run real solar data.

\section{RESULTS}
\subsection{Overall description of the active region}
NOAA AR 12371 displayed a complex magnetic field configuration (Figure~\ref{hmi_full}), with its leading sunspot consisting of a single negative magnetic polarity, and the trailing part consisting of multiple spots with mixed magnetic polarities that resulted from emergence of satellite polarities inside the main bipole. These trailing spots share the same penumbral structure (upper-right panel), making it a $\delta$-sunspot. NOAA AR 12371 was located at the polarity inversion line of the background field. This made the negative polarity of the AR be the parasitic polarity situated inside the surrounding positive polarity. Such magnetic configuration is favorable for the so called fan-spine topology of the coronal magnetic field.

SDO/AIA coronal images show presence of an inverse-S sigmoid structure (lower-left panel) inside the AR. Note that the existence of a sigmoid in an AR is considered to be a precursor of a possible eruption. The core of this sigmoid, seen as dark threads in the AIA~193~\AA\ image, connects opposite polarities of the $\delta$-sunspot and is observed as a compact, bright thread in the Hinode/XRT image (lower-right panel). The outer part of the sigmoid is less intense and relatively diffused. Also, it seems to be composed of two sets of J-shaped loops. The loops in the lower part of the sigmoid are connecting the positive polarity of the $\delta$-spot, P1, and the negative polarity of the leading spot, N1. This indicates the field connectivity between P1 and N1 is on a larger scale compared to the main sigmoid field connecting P1 and N2. Those loops are also visible in the AIA~193~\AA\ image.

\subsection{Magnetic field structure of the active region and its neighborhood background}
\begin{figure}[tb]
\begin{center}
    \includegraphics[width=8cm]{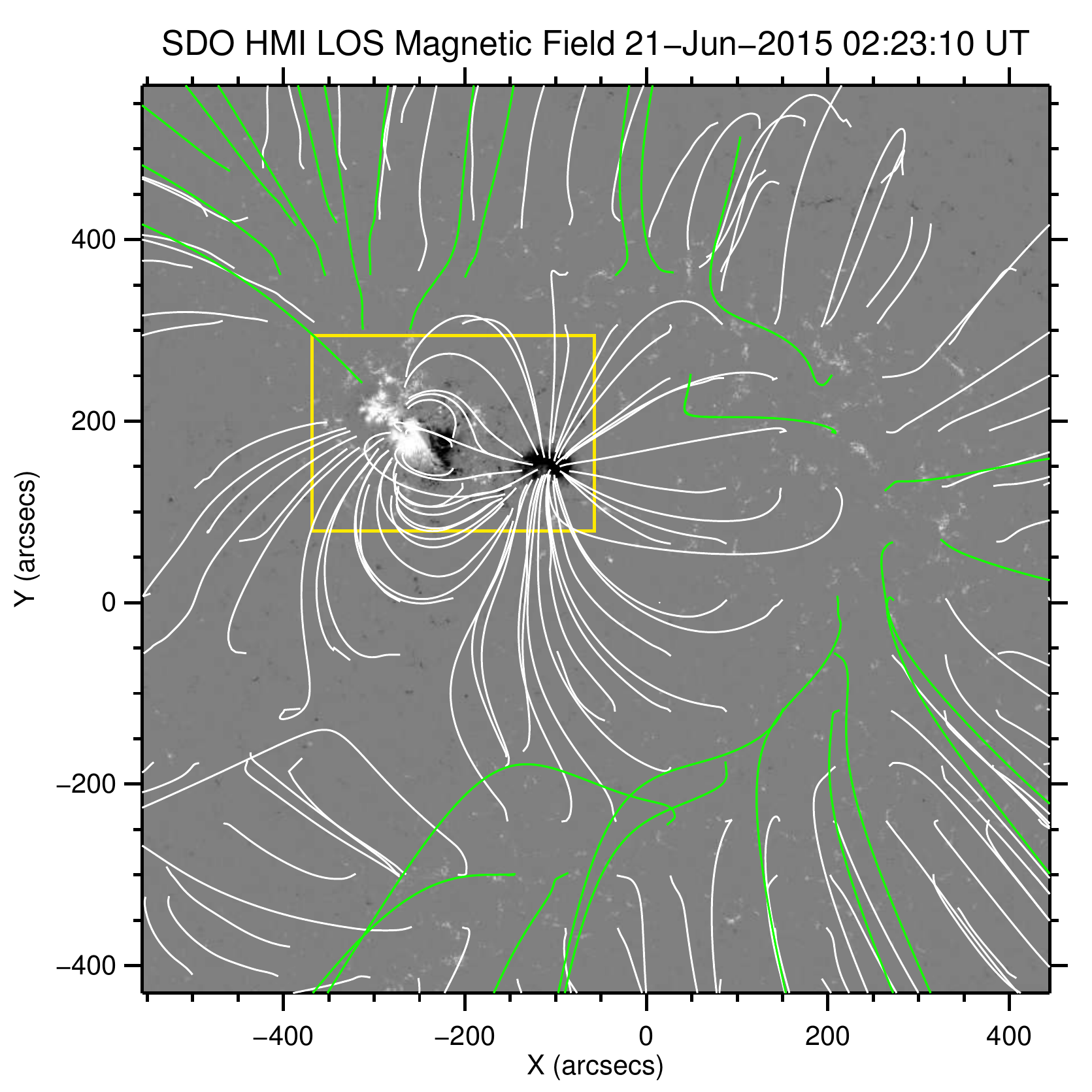}
    \includegraphics[width=8cm]{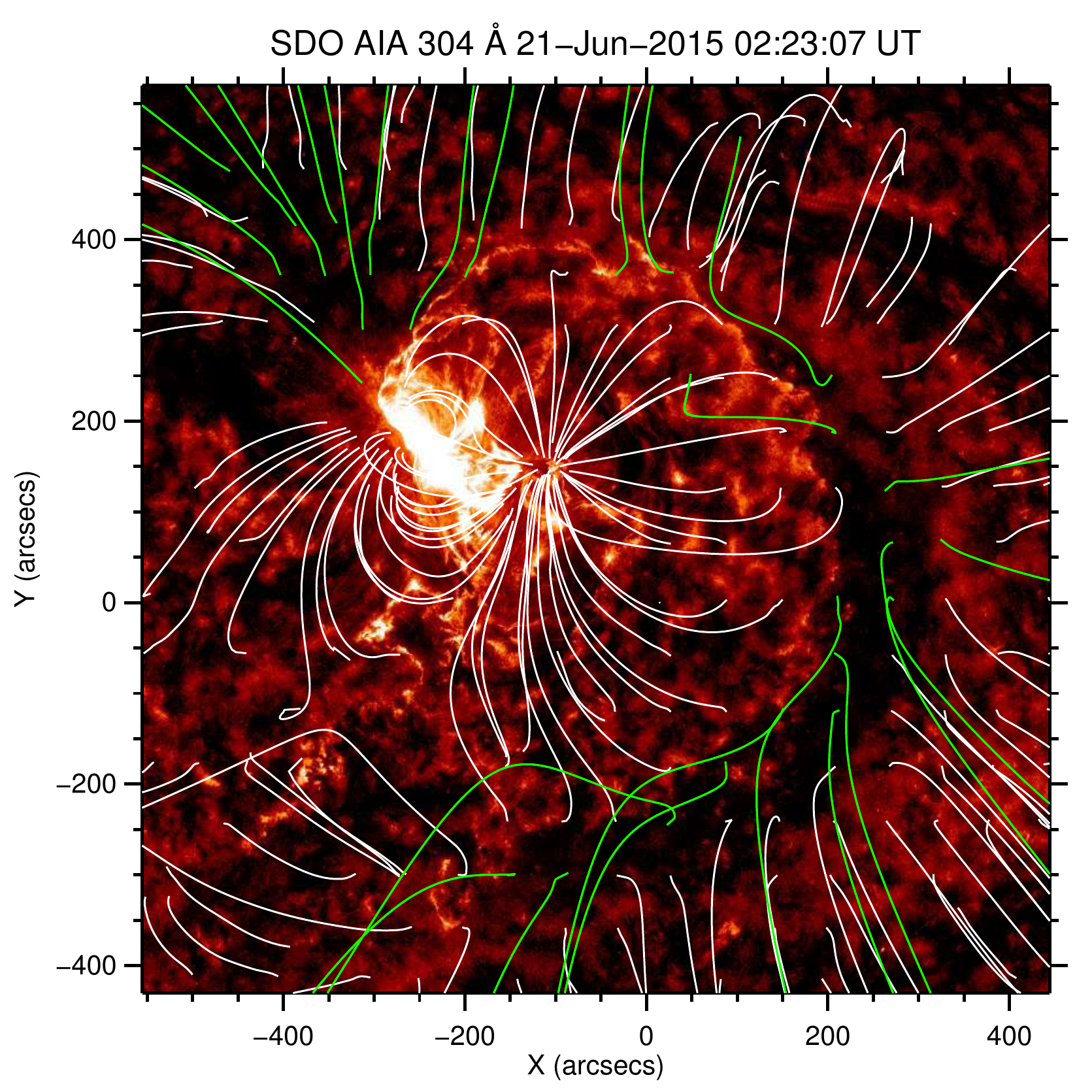}\\
    \includegraphics[width=8cm]{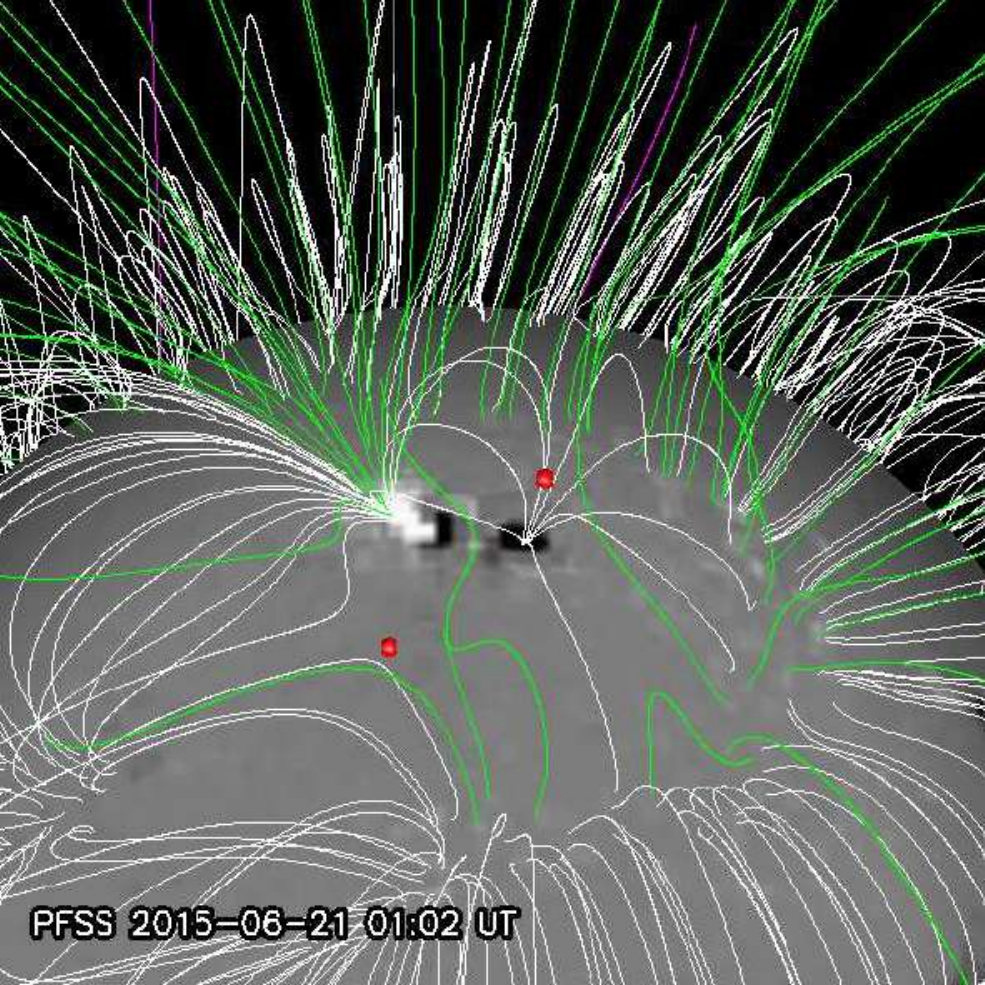}\,\,
    \includegraphics[width=8cm]{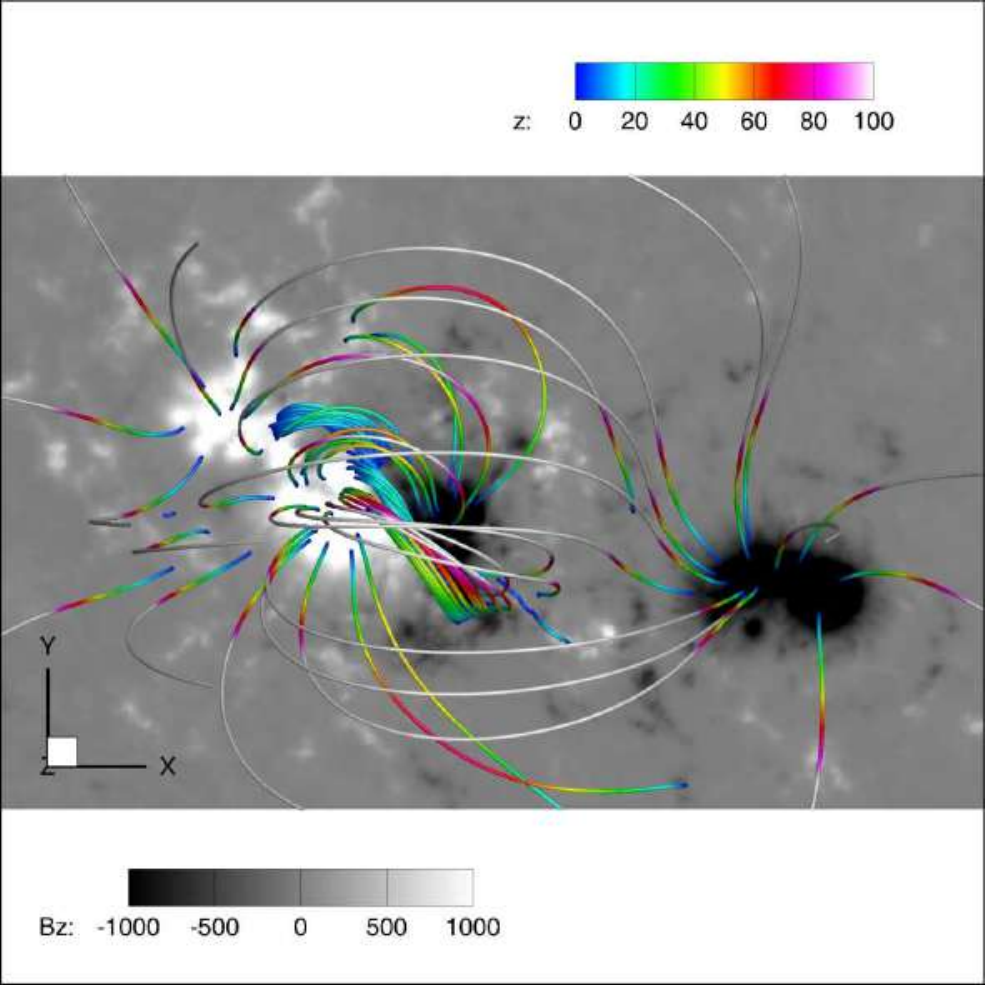}
    \caption{Coronal magnetic field extrapolation results for NOAA AR 12371. Some selected magnetic field lines from PFSS model are overplotted on the photospheric magnetogram (upper-left, lower-left) and AIA~304\AA\, data (upper-right). Closed field lines are colored in white and open field lines are colored in green. Null points calculated from the PFSS magnetic field are presented by red dots (lower-left). Magnetic field lines obtained from NLFFF model within the FOV represented by a yellow box in the upper-left panel are displayed in the lower-right panel with the photospheric magnetogram. Height scale along the field lines represented by different colors are specified as a color bar.}\label{pfss}
\end{center}
\end{figure}

It is worth examining the underlying magnetic field topology to understand the consecutive activities that ensued in this AR. The upper and lower-left panels of Figure~\ref{pfss} show the PFSS model plotted over an HMI magnetogram (left panels) and the AIA~304~\AA\, data (upper-right panel). Closed field lines are plotted in white color and open field lines are in green. As it follows from the PFSS model, the magnetic field lines rooted at N2 in general are of a fan-spine shape with the inner-spine anchored at N2. The field lines fanning out eastward are mostly connected to P1 and those fanning out westward are mainly rooted at dispersed background field. The fan-shaped field lines are bounded by open field lines (green) at the northern and western side (upper-right panel). Some of these open field lines are bending inward and converge to form an outer-spine-like structure. The footpoint of the outer spine is not distinguished in this model, and possibly due to the large scale of the studied system so that it may be located beyond the FOV, or the outer spine may be an open field line. Even with only potential field extrapolation, we could see that the fan-shaped magnetic topology may have played a role in forming a distinctively shaped QCSR.

The PFSS model also shows a small magnetic arcade below the fan-shaped structure that connects opposite polarities of the $\delta$-spot. This arcade was reconstructed as a twisted flux rope structure in the NLFFF model (lower-right panel), whose twist direction and the overall shape are consistent with the observed sigmoid. As shown in the figure, the low lying, highly sheared magnetic field structure is nearly parallel to the polarity inversion line of the $\delta$-spot, and is enveloped by highly extended magnetic field lines that correspond to the fan-shaped structure of the PFSS model.

\subsection{Overall flare activities}
\begin{figure}[tb]
\begin{center}
    \includegraphics[width=9cm]{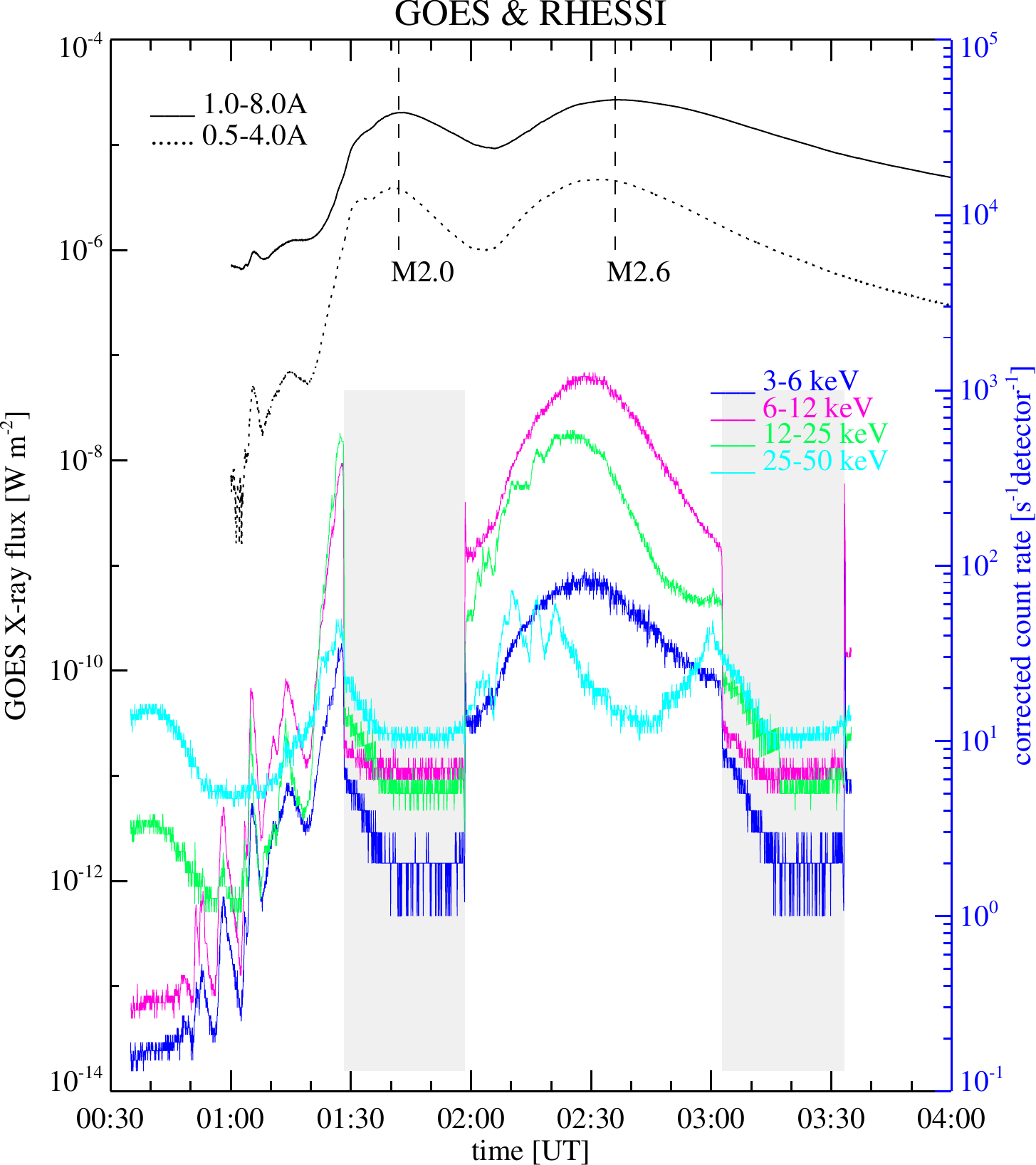}
    \includegraphics[width=10cm]{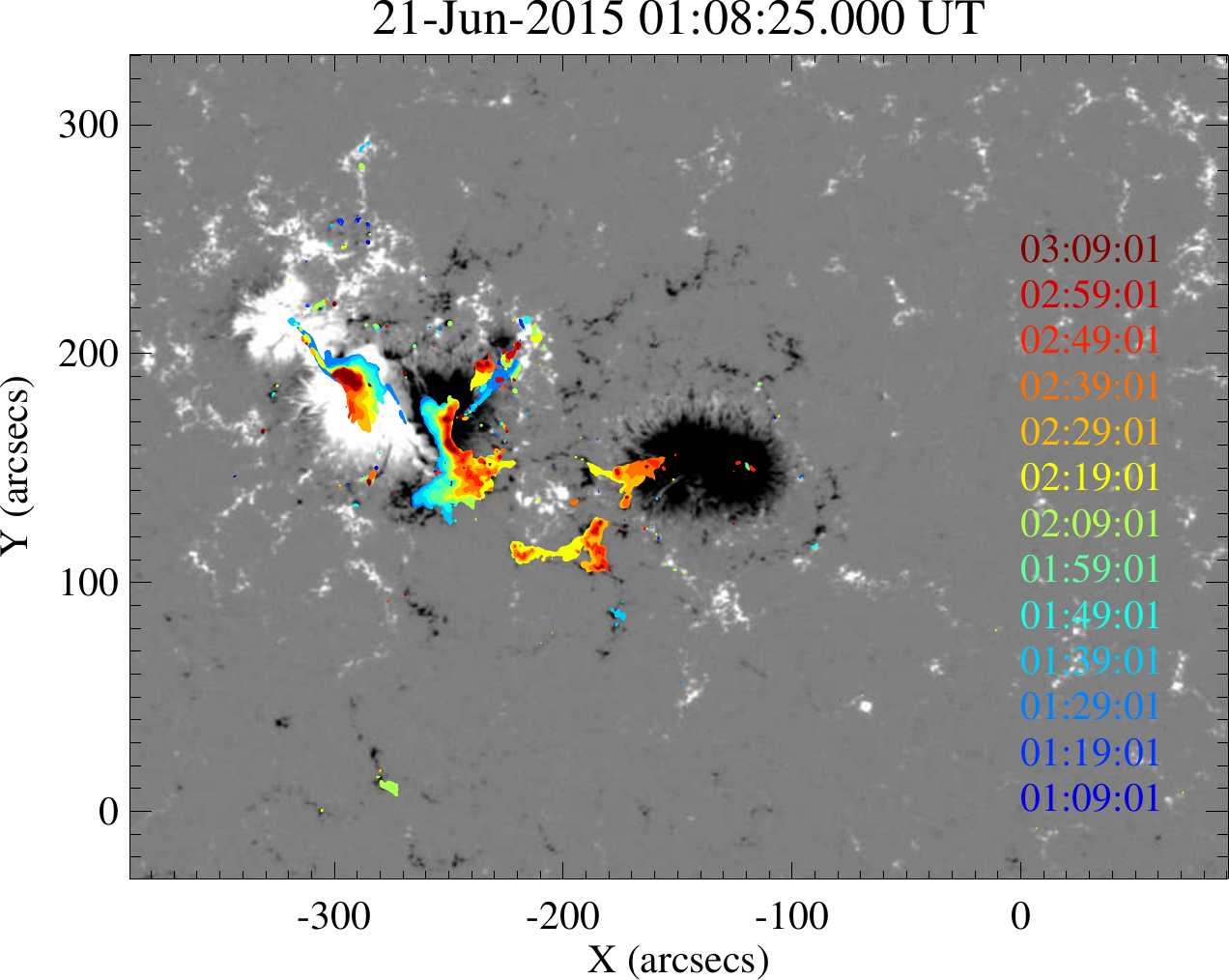}
    \caption{Top : \emph{GOES} X-ray flux (upper black curves) and RHESSI count rate (lower colored curves) during the observed period. Peak times of the two M-class flares, M2.0 at 01:42:00UT and M2.6 at 02:34:00 UT, are indicated with dashed lines. Two shaded time regions show the night time of RHESSI satellite. Bottom : Temporal changes of the main parallel ribbon defined using AIA~1600\AA\ data.}\label{goes}
\end{center}
\end{figure}

According to the \emph{GOES} X-ray flux plot (Figure~\ref{goes}), two M-class flares successively occurred on 2015 June 21 with a 54-minute time interval. Flare II started before the intensity of Flare I dropped to its pre-flare value, and its decay phase lasted for over two hours, making it a long duration event. In the RHESSI plot, despite the data gap during Flare I due to the satellite night time, one can detect multiple energy release events in both thermal (blue) and non-thermal (green and cyan) components during the pre-flare phase. Similar multiple non-thermal peaks were also detected at the early phase of Flare II, which may be indicative of pre-flare activities in which multiple magnetic reconnections such as tether-cutting are likely to be involved \citep{Yur15, Kum17}.

\begin{figure}[tb]
\begin{center}
    \includegraphics[width=1\textwidth]{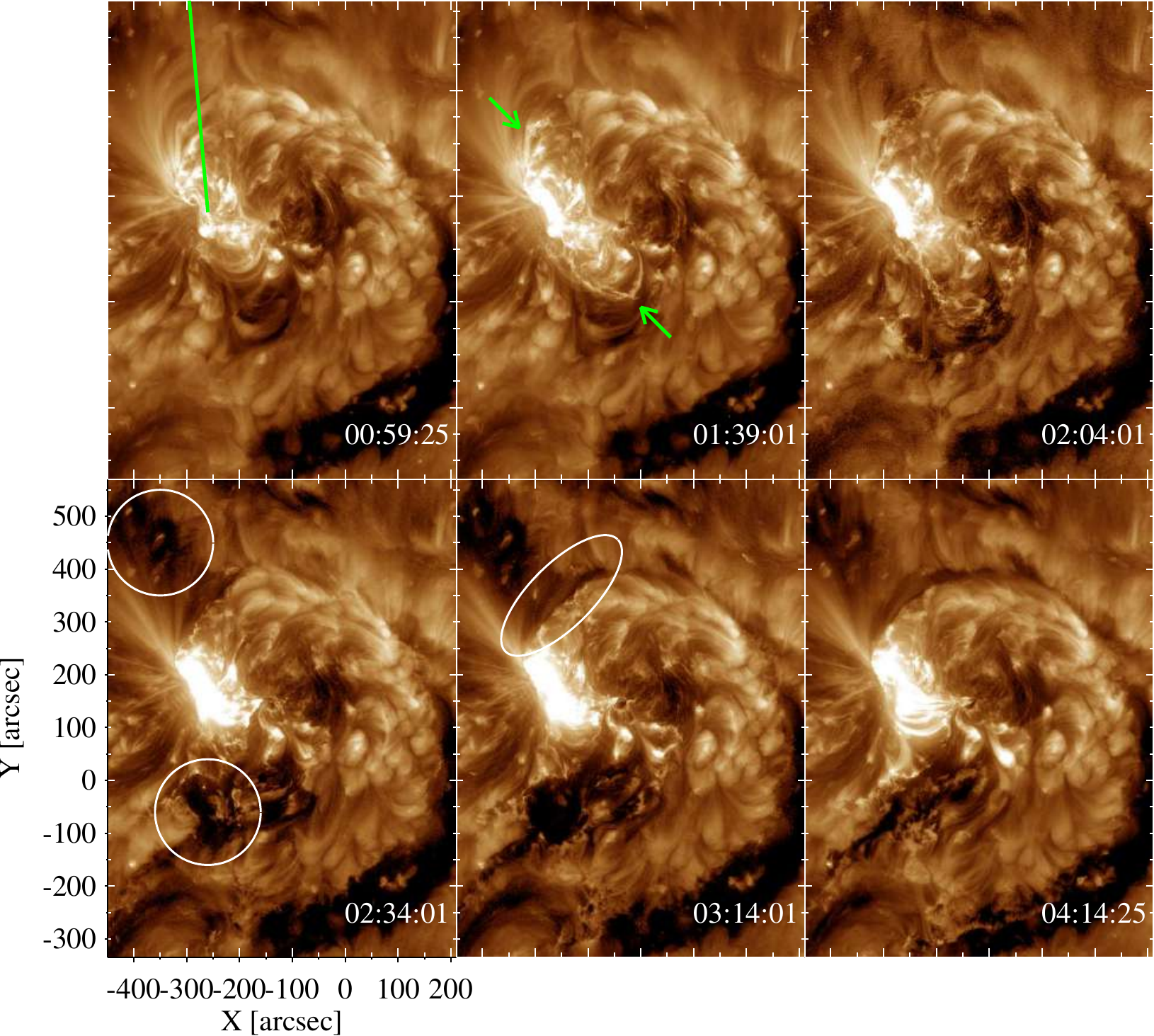}
    \caption{AIA~193\,\AA\, images showing the temporal variation of coronal structures during Flare I and II. Expanding coronal loops initiated at Flare I are indicated by green arrows, and locations of both main dimmings and arc-shaped dimming are presented by white circles and ellipse, respectively.}
\label{flare_aia193_intensity}
\end{center}
\end{figure}

Overall EUV evolution of the successive flares is presented in Figure~\ref{flare_aia193_intensity}. Sequential AIA 193~\AA\ images cover about three hours including the onset and the peak times of each flare as well as the decay phase of Flare II. The main brightening in both flares appeared at the sigmoid structure (01:39:01~UT, 02:34:01~UT), while the bright postflare loops in case of Flare II were more extended toward N2 (02:34:01~UT). This indicates that the sigmoid is the region where the flare magnetic reconnection takes place, and magnetic field involved in Flare II is of a larger scale than that involved in Flare I. Separation of the two-ribbons observed in AIA~1600\AA\ (Figure~\ref{goes}, bottom panel) also supports this interpretation. The color shaded area represents the ribbon area at each time interval, with the blue shade showing ribbons at earlier times and the red at later times. In addition to the typical ribbon separation, there is an interesting behavior such as sudden appearance of a third flare kernel near N2 at 02:19~UT close to the beginning of Flare II. This fact indicates that the magnetic field connecting N2 and P1 that is a part of the fan-shaped structure, went through reconnection during Flare II as well, in addition to the magnetic field connecting N1 and P1, that is a part of the sigmoid.

\begin{figure}[tb]
\begin{center}
    \includegraphics[width=0.5\textwidth]{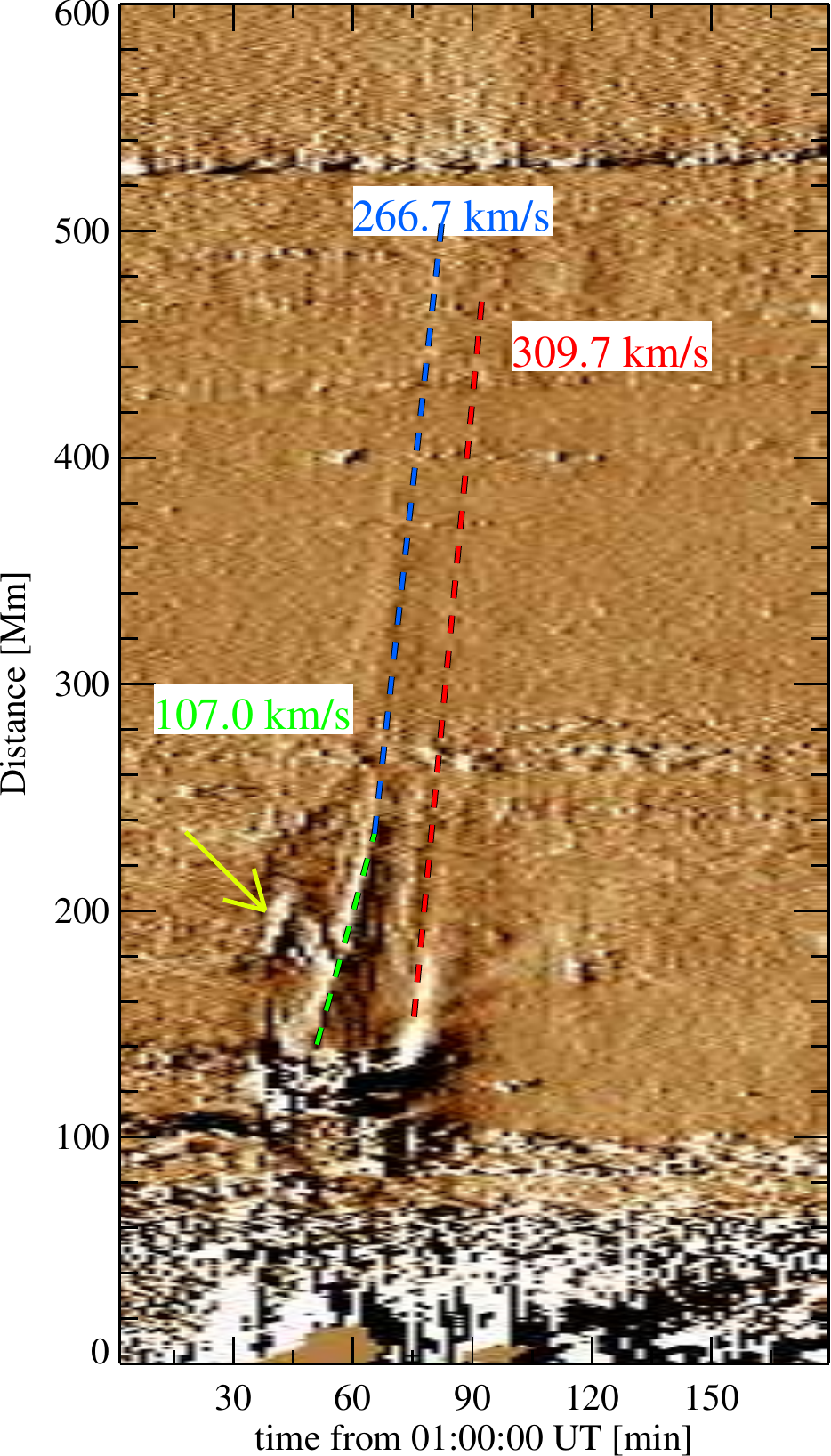}
    \caption{Space-time diagram obtained along a slit shown in Figure~\ref{flare_aia193_intensity}. Two tracks of expanding coronal loops in the northern part of the AR are distinguished. One track started at the time of Flare I with the projected speed of \kms{107.0} (green) then accelerated to \kms{266.7} after Flare II (blue). The other track of ejecta with the speed of \kms{309.7} (red) started at the onset time of Flare II. A concave-up shaped track indicated by an yellow arrow is also detected indicating some loops contracted after a short expansion. }\label{x-t diagram}
\end{center}
\end{figure}

During Flare I, we detected expanding bright flare loops in both northern and southern part of the sigmoid in AIA~193~\AA\ images (green arrows in 01:39:01~UT panel of Figure~\ref{flare_aia193_intensity}). The projected speed of this expansion estimated along the artificial slit shown in Figure~\ref{flare_aia193_intensity} was found to be \kms{170} until $\sim$ 02:07~UT and then it suddenly increased to \kms{267} just after Flare II onset (Figure~\ref{x-t diagram}). Simultaneously with this acceleration, another track of loop expansion indicated by the red dashed line was initiated with the projected speed of \kms{310}. Based on the abrupt change of the expansion speed near the start of Flare II and the following expansion with the higher speed, it seems that either weakening or removal of the fan-shaped field envelope may have occurred with Flare II initiation. The AIA~193~\AA\ image at 02:34:01~UT (Figure~\ref{flare_aia193_intensity}) shows that loops in the northern part of the sigmoid are now more straightened and similar to the open field, indicating that loops have significantly expanded. In addition, two well defined dimmings developed at both endpoints of the sigmoid, suggesting density depletions due to mass loss in the regions. Dimmings are considered to be an evidence of a CME eruption, and indeed, a halo CME was detected by the LASCO/C2 coronagraph on June 21 at 02:36:05~UT (more details in Section~\ref{section_cme}). In addition to the two main dimmings, intensity in the region swept by an additional ribbon brightening (inside a narrow ellipse) has also notably decreased to form an arc-shaped dimming (02:34:01~UT, 03:14:01~UT), which even slightly increased as Flare II evolved (04:14:25~UT), while the area of the main dimming decreased.

A sequence of AIA~304~\AA\ images in Figure~\ref{flare_aia304} show similar flare development, along with an additional insight into the flare ribbon structure. Although the two-ribbon structure is not well distinguished here due to the overall brightness of the flare, sharp edges of the bright post-eruption arcade (PEA) highlighted by green lines (01:56:49~UT, 02:59:49~UT) gives us a hint of ribbon location. Those edges show separation and were connected by the PEA as typical flare ribbons do. In addition to the two regular ribbons, a secondary ribbon can also be well detected in the data, for instance at 02:59:49~UT and 03:59:49~UT. This ribbon had a quasi-circular shape, but it did not fully enclose the southern part of the AR, and it was strikingly large in size spanning nearly 650\arcsec\ in north-south direction and 550\arcsec\ in east-west direction. The secondary ribbon structure was less bright than the regular ribbons and was not connected by any PEA. These characteristics are similar to those reported by \citet{Zha14}, except the ribbon life time was much longer in our case. This QCSR first appeared at the late phase of Flare I (01:56:49~UT) north of the flaring sigmoid, then it sequentially brightened and extended encircling the flaring AR in the clockwise direction as Flare II evolved. Southern part of the QCSR was observed even after 04:00~UT.

\subsection{Quasi-circular secondary ribbon}
\begin{figure}[tb]
\begin{center}
    \includegraphics[width=1\textwidth]{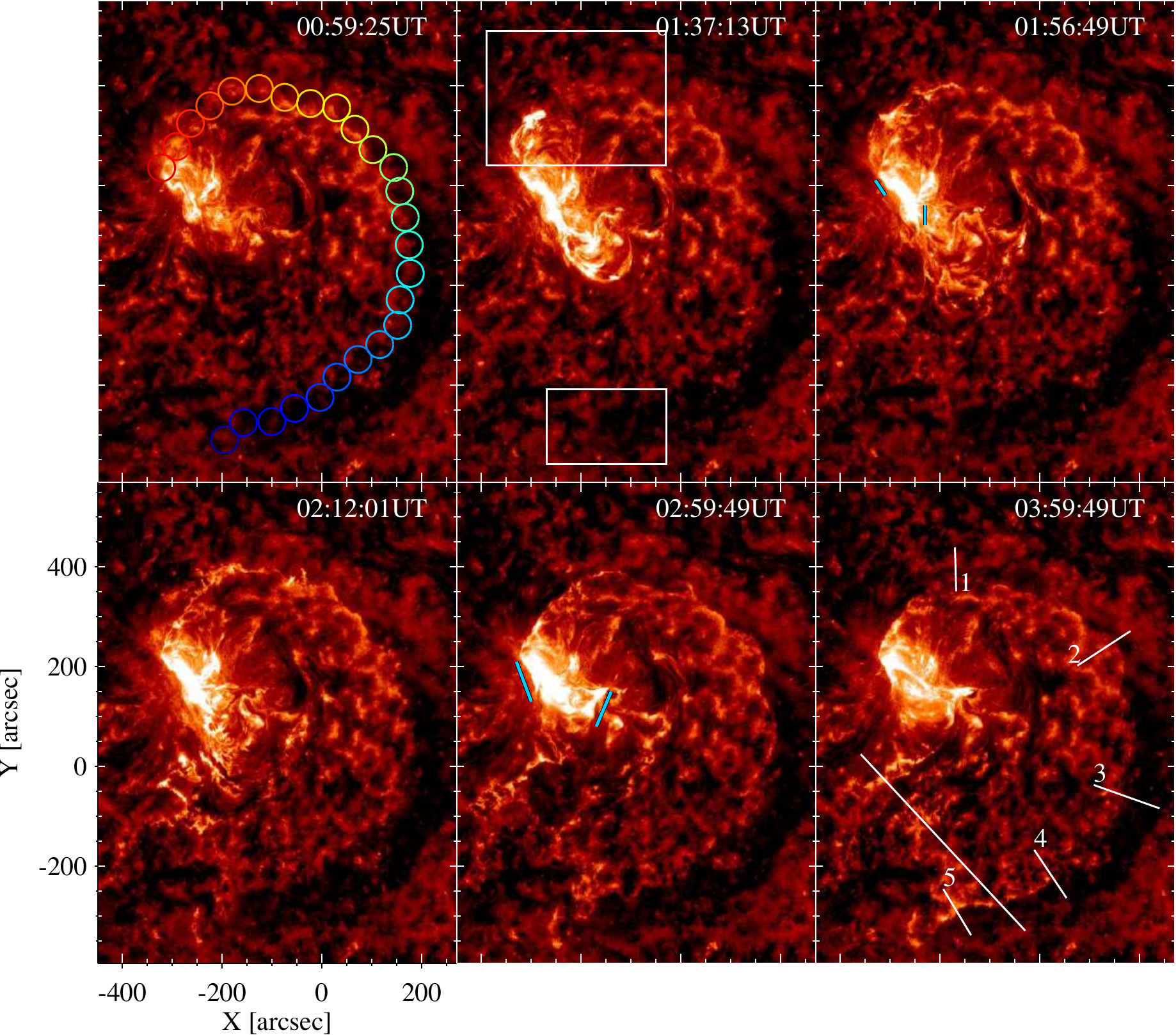}
    \caption{AIA~304~\AA\ images showing the temporal development of the QCSR (indicated by series of colored circles) during Flare I and II. The FOV is the same as that of Figure~\ref{flare_aia193_intensity}. Centers of colored circles are manually defined along the ribbon spine at its best visibility, and a diameter of each circle is 60\arcsec\,. Cyan lines outline the locations of two parallel flare ribbons, and white lines are representing artificial slits along which space-time diagrams shown in Figure~\ref{qcsr_xt_normal} are obtained. White boxes represent regions where fine structures of the QCSR will be studied.} 
\label{flare_aia304}
\end{center}
\end{figure}

\begin{figure}[tb]
\begin{center}
    \includegraphics[width=0.4\textwidth]{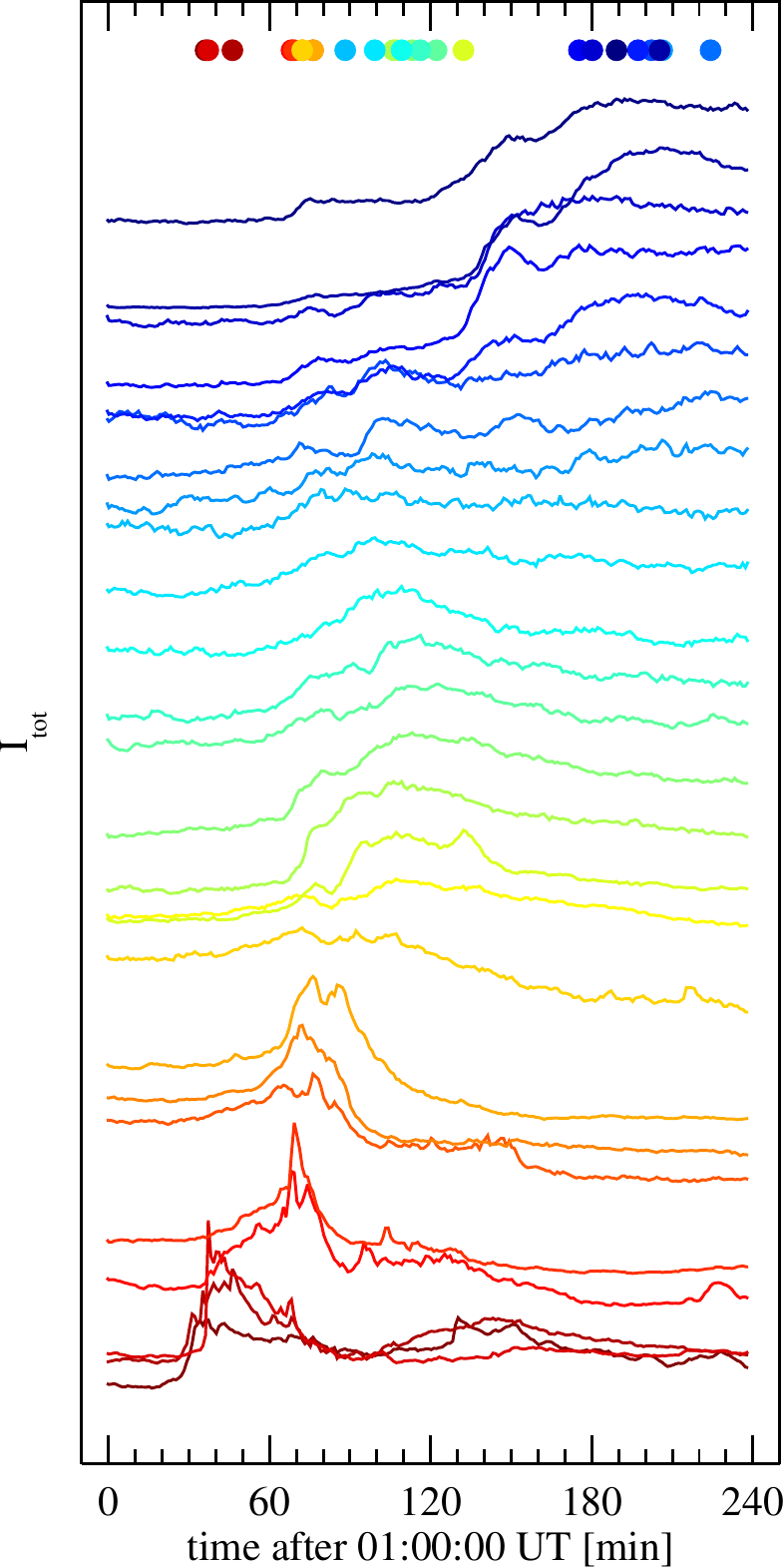}\,
    \includegraphics[width=0.5\textwidth]{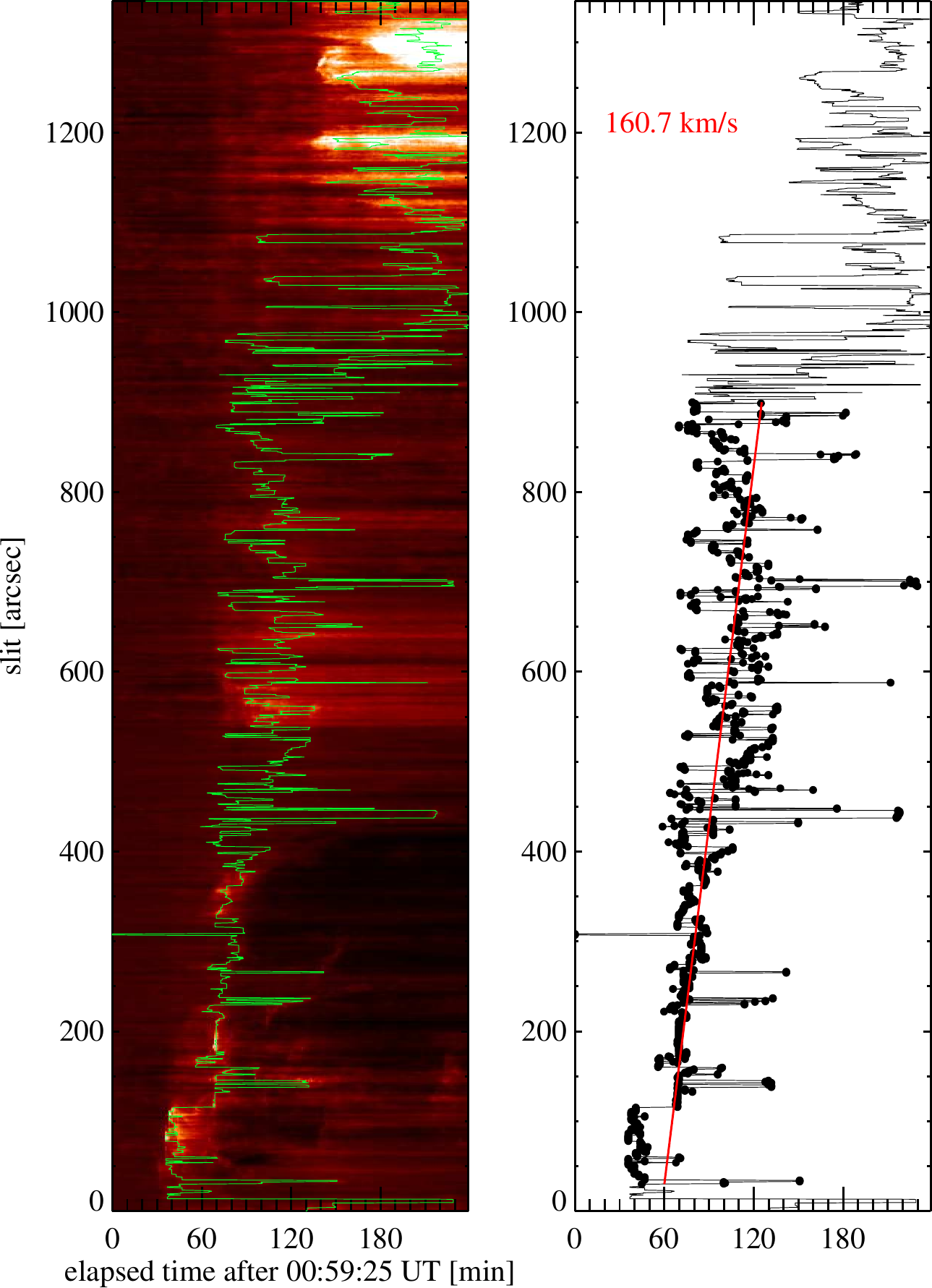}
    \caption{Left: light curves of the QCSR measured within each circles shown in Figure~\ref{flare_aia304}. Peak time of each light curve is indicated by a closed circle with the same color-code. Middle: space-time diagram obtained from the curved slit running along the QCSR spine. Intensity peak time at each pixel along the slit was calculated and overplotted with green color. Right: The same intensity-peak-time plot shown in the middle panel. The result of a linear fit to the intensity peak times to estimate the apparent speed of the subsequent ribbon brightening along the QCSR, \kms{161}, is presented by a red line. }\label{qcsr_xt_along}
\end{center}
\end{figure}

In order to understand dynamical properties of the observed QCSR, we measured the temporal change of the ribbon intensity both along and across the ribbon. We first defined the spine of the curved ribbon by manually choosing 26 points along the bright ribbon edge at the time of its best visibility. Then we defined 26 circles with a radius of 27\arcsec\ centered at each point as plotted in the first panel of Figure~\ref{flare_aia304}. Light curves of the total intensity measured within each circle are presented in the left panel of Figure~\ref{qcsr_xt_along} with the same color-coding we used to plot the circles in Figure~\ref{flare_aia304}. The peak time of each light curve is indicated by filled circles of the same color as the associated light curve. There is a tendency that the part of the QCSR closer to the flare site earlier in time. Also, the shape of each light curve varies as the sampling location moves farther away from the flaring AR in the clockwise direction. For instance, the red color light curve group representing the part of the QCSR closer to the flaring AR shows an impulsive rise of intensity during about ten minutes and then the intensity falls off for about 30 minutes, while the light curves in the blue color group measured farther from the flare show a long-duration-like pattern.

The apparent speed of the subsequent brightening along the QCSR was estimated by obtaining a space-time (x-t) diagram from the curved slit along the QCSR spine. We defined a curved slit by interpolating the circle center points, and then set the width of the slit to be 60\arcsec\ wide enough to account for expansion of the QCSR normal to the spine. The total ribbon intensity within the width of the slit was measured in time and the resulting x-t diagram is displayed in the two right panels of Figure~\ref{qcsr_xt_along}. The slit length increases clockwise starting from the center of the first dark red circle closest to the flaring AR. The middle panel is a normal x-t diagram, while the intensity peak time at each pixel along the slit is plotted by the black closed circles in the right panel. We performed linear fit to the intensity peak curve and estimated the apparent speed of the subsequent brightening along the QCSR to be \kms{160.7}. This value only slightly exceeds the \kms{140} value derived by \citet{Zha14} from AIA~304~\AA\ data.

\begin{figure}[tb]
\begin{center}
    \includegraphics[width=0.45\textwidth]{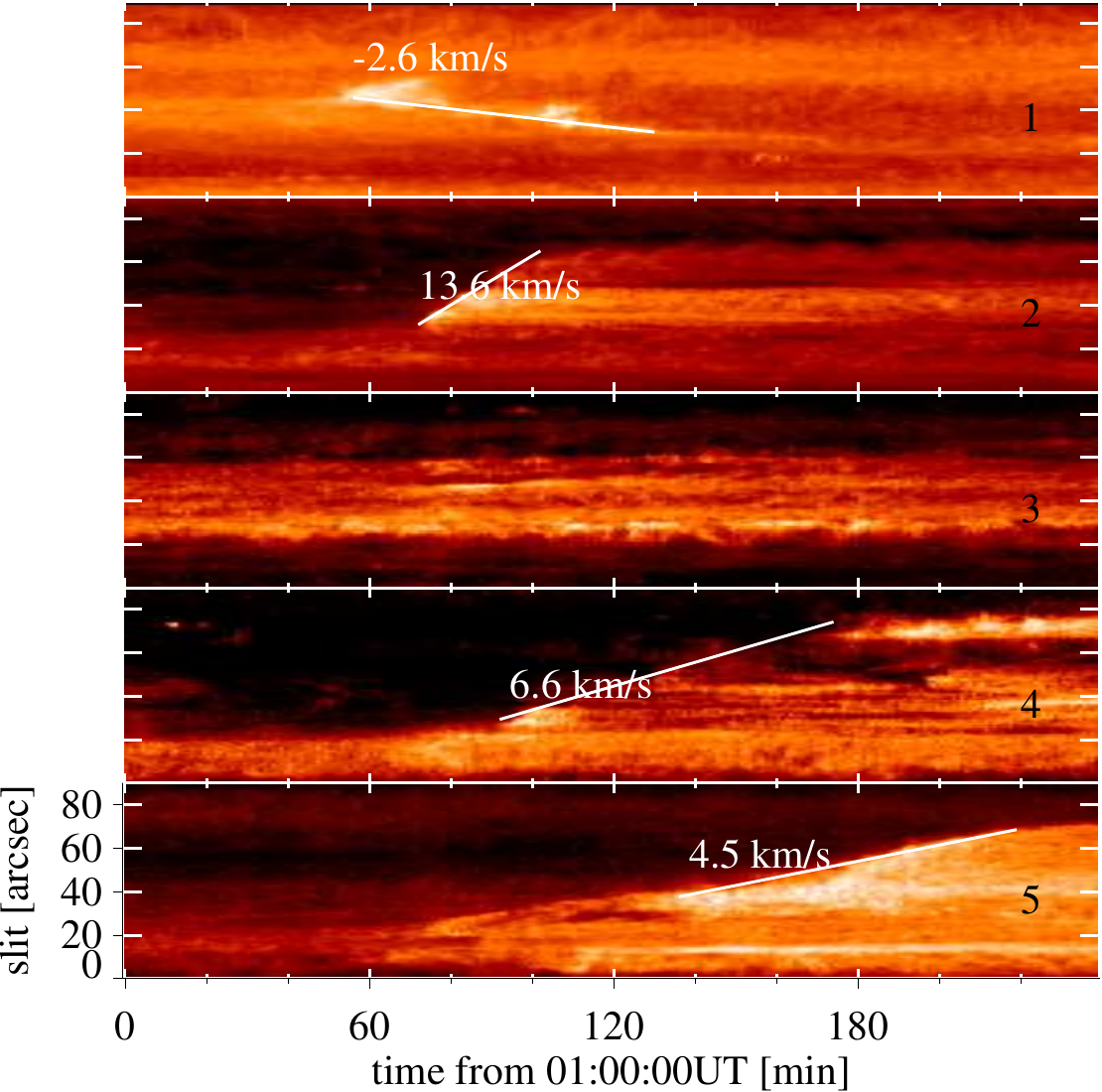}\,
    \includegraphics[width=0.45\textwidth]{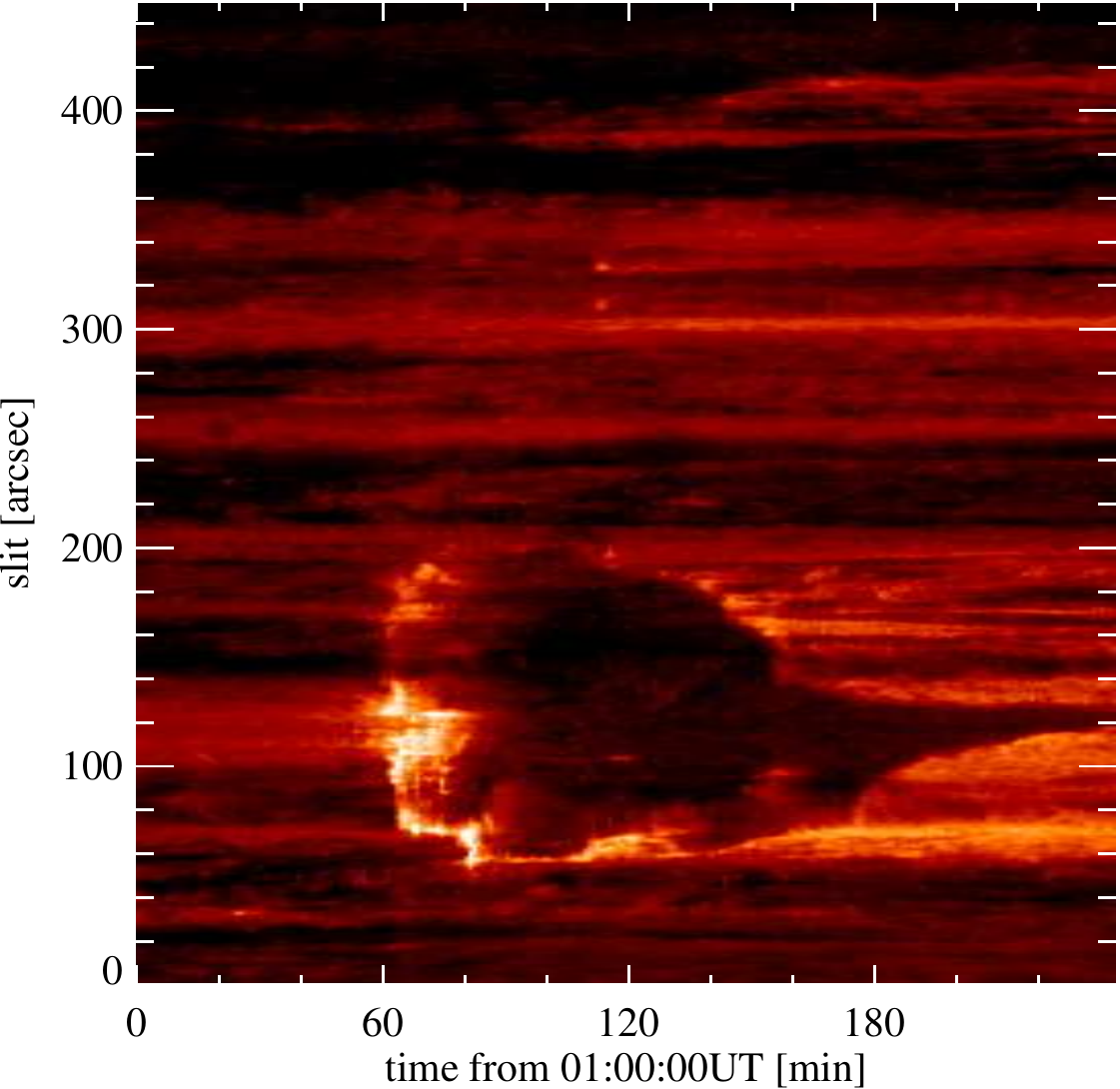}
    \caption{Space-time (x-t) diagrams obtained along the short slits numbered from 1 to 5 shown in Figire~\ref{flare_aia304} (left) and along the longest slit (right). The distance in each x-t diagram increases in the direction away from the center of the QCSR. The estimated speed of ribbon expansion or contraction along the slit direction is presented with the resulting linear fit (white line). }\label{qcsr_xt_normal}
\end{center}
\end{figure}

We already mentioned that the observed QCSR not only showed subsequent brightenings along the ribbon spine but also normal to the spine, in other words, radial contraction or expansion of the QCSR. We constructed x-t diagrams along a number of slits normal to the ribbon spine shown in the last image of Figure~\ref{flare_aia304}. The left panel of Figure~\ref{qcsr_xt_normal} shows five x-t diagrams obtained from the short slits numbered from 1 to 5 in Figure~\ref{flare_aia304}, and the right panel shows a x-t diagram along the longest slit. The distance in each x-t diagram increases in the direction away from the center of the QCSR. The five arbitrary slit positions were chosen to highlight various behavior of radial expansion. Both speed and duration of the ribbon expansion vary depending on the position along the circular ribbon. At slit position 2, the bright part of the ribbon expanded by approximately 40\arcsec\ in 30~min yielding the apparent speed of \kms{13.6}, while no expansion was detected at slit 3 and a contraction was detected at slit 1. Clear ribbon expansion was detected mostly in the southern part of the ribbon, and only a small fraction of the north part of the ribbon close to the flaring AR showed contraction. Such contracting or expanding motions of a QCSR has not yet been reported. \citet{Zha14} reported migration of the magnetic field normal to the ribbon axis, and in that case the derived speed was much lower ($\sim$\kms{0.4}) than the speeds of ribbon contraction/expansion found here. Moreover, a weak dimming following the contraction of the bright ribbon edge was detected in the x-t diagram at slit 1. This is the location where an arc-shaped dimming was observed (see Figure~\ref{flare_aia193_intensity}). This observation implies the ribbon contraction at slit 1 may be related to field opening followed by plasma ejection.

The right panel of Figure~\ref{qcsr_xt_normal} shows an x-t diagram along the long slit shown in Figure~\ref{flare_aia304}. As we mentioned above, we observed two types of coronal dimmings in AIA~171~\AA\ and AIA~304\AA\, images formed just after Flare II : the main double dimmings near the footpoints of the sigmoid and an arc-shaped narrow dimming just outside the northern bright edge of the QCSR. To examine their temporal variations, we positioned the long slit cross the southern main dimming. The dimming onset was preceded by an intense brightening at around 02:02~UT, which is near the start time of Flare II. The dimming then rapidly reached its maximum size at around 02:35~UT, which is near the peak time of Flare II. These data strongly supports our conjecture that the main plasma ejection (halo CME) was initiated in relation to Flare II. The main dimming showed a relatively quick recovery between 02:35~UT and 03:40~UT, followed by a gradual recovery until around 05:00~UT. It is worth pointing out that the recovery time of 2.3 hours is much shorter than the mean recovery time of 6 hours that was previously reported in a statistical study by \citet{Kri17}.

\begin{figure}[tb]
\begin{center}
    \includegraphics[width=1\textwidth]{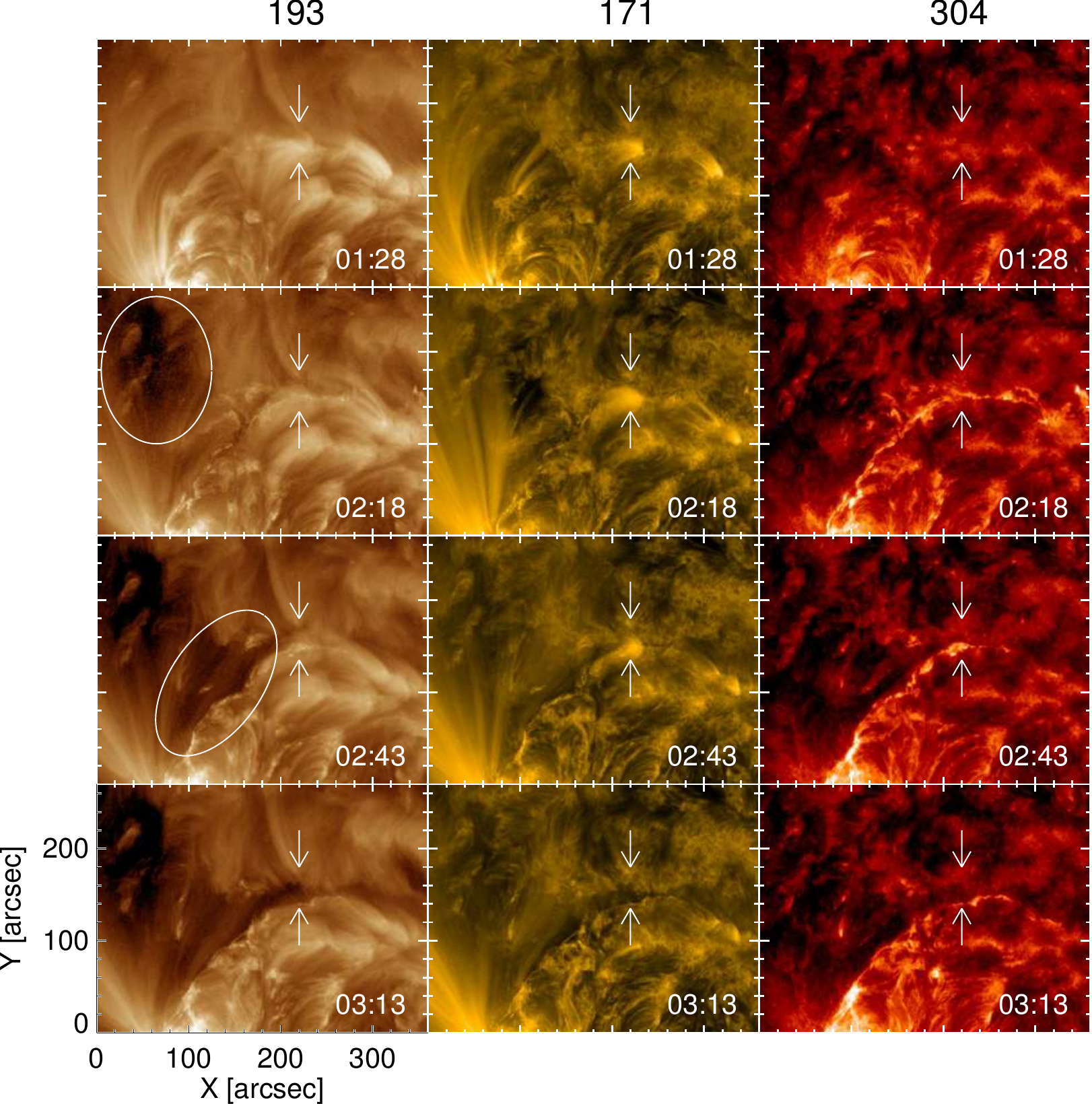}
    \caption{AIA image sequence within the northern white box shown in Figure~\ref{flare_aia304} at different time instances. The spatial range swept by the QCSR brightening is outlined by two white arrows, and locations of main and arc-shaped dimmings are presented by white ellipses. }\label{aia_3chan_north}
\end{center}
\end{figure}

The boxes in the upper-middle panel of Figure~\ref{flare_aia304} enclose two different parts of the QCSR that showed either contraction (top box) or expansion (bottom box). The time evolution of the northern part of the QCSR is shown in Figure~\ref{aia_3chan_north}. Among the three selected channels, the bright ribbon is best visible in the AIA~304~\AA\, images, while the coronal loops and dimmings are well detected in the 193~\AA\, images. The earliest AIA~193~\AA\, image was taken before the onset of Flare II and shows long and arch shaped AR coronal loops and short, and bright patches that seem to be footpoints of quietsun region loops. These short patches are best distinguished in AIA~171~\AA\, images. At the time of QCSR appearance (02:18~UT), a thin and bright ribbon structure (which was identified as the same QCSR) appears in AIA~304~\AA\, image (between two white arrows). A similar structure is also visible in other AIA channels. This is the time when the main dimming begins to form (vertical ellipse). In the subsequent images, the bright ribbon continuously shifted southward (AIA~304~\AA\,), leaving its previous location dimmed (AIA~193~\AA,, inclined ellipse). Note that the arc-shaped dimming following the ribbon contraction is not localized at the sigmoid footpoint, but is extended along the outer boundary of the northern QCSR.

\begin{figure}[tb]
\begin{center}
    \includegraphics[width=1\textwidth]{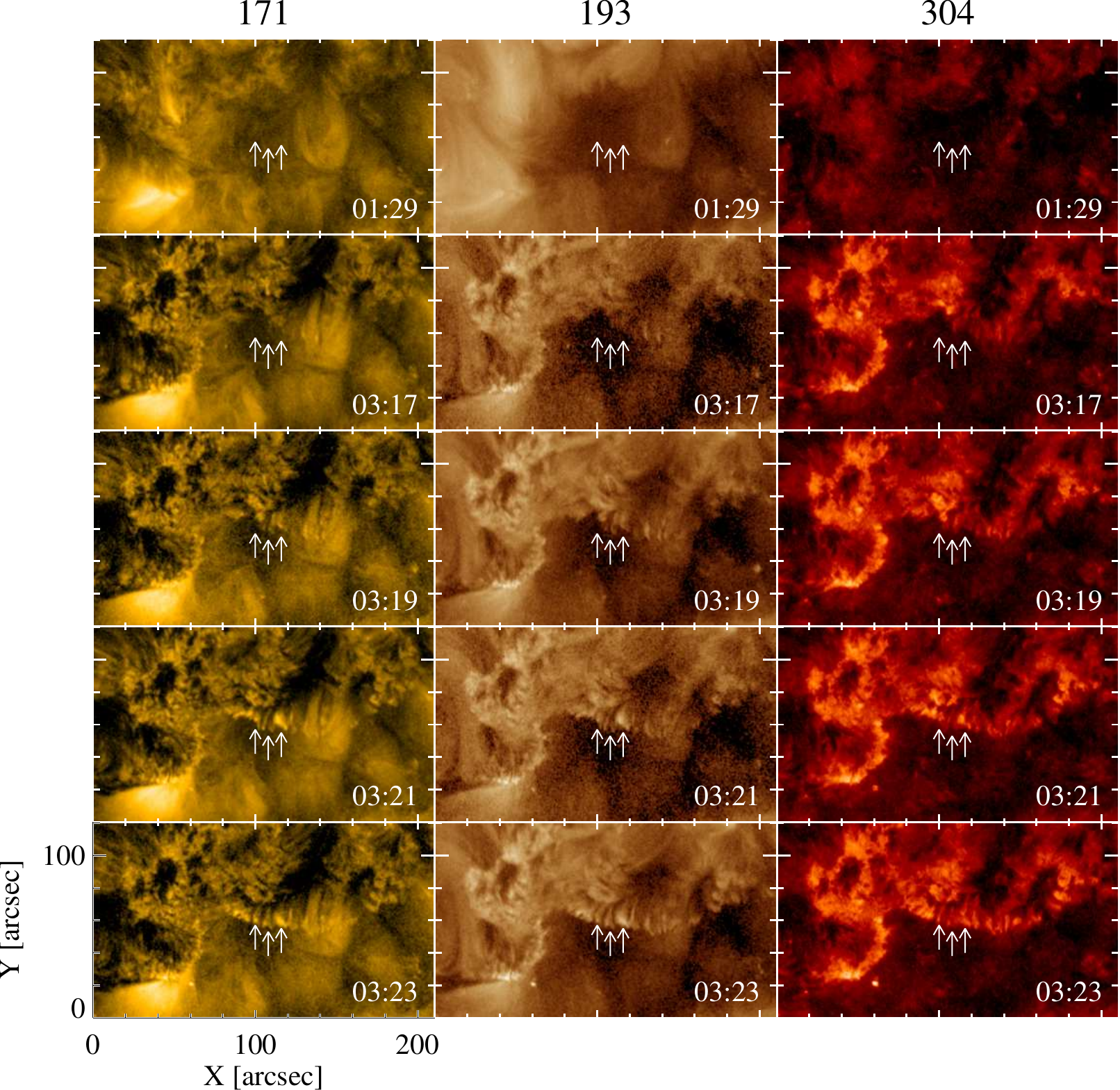}
    \caption{AIA images within the southern white box shown in Figure~\ref{flare_aia304} at different time instances. Fine structures of the QCSR with a finger-tip shape are pointed by white arrows.  }\label{aia_3chan_south}
\end{center}
\end{figure}

Figure~\ref{aia_3chan_south} shows the southern part of the QCSR. The bright ribbon in this region first appeared at around 3:00~UT, and then it expanded southward. Before the QCSR appearance and they become sharp and well defined, structures in the AIA~171~\AA\, image at 01:29~UT were seen smudged and diffused. This may be another signature of plasma ejection. While both main and arc-shaped dimmings are limited to certain areas, the contrast enhancement was observed over a large area within the QCSR region suggesting that plasma ejection occurs not only within the footpoints of the sigmoid but over a larger area occupied by the fields involved in a flare. Another finding is the finger-tip shaped fine structures, which are persistent and retain both their shape and the initial locations as the QCSR expands farther southward (pointed by three arrows). Each finger-tip structure was oriented normal to the outer edge of the ribbon, and their length increased as the ribbon expanded. The distance between finger-tip structures is estimated to be around 6\arcsec\,, and their cross section is of order of 1\arcsec\, -- 2\arcsec\,.

\subsection{Halo CME associated with flares}\label{section_cme}
\begin{figure}[tb]
\begin{center}
    \includegraphics[width=0.8\textwidth]{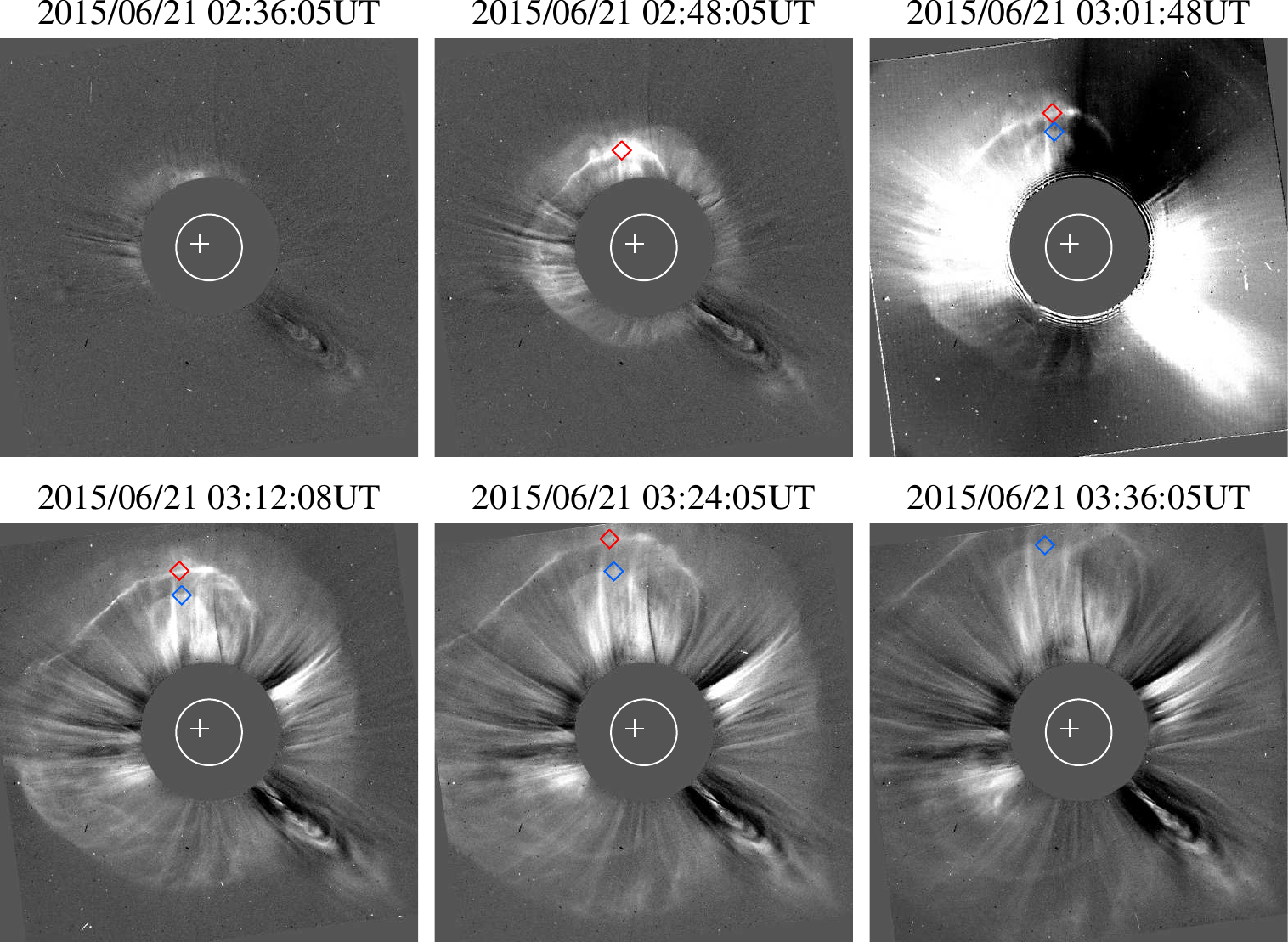}
     \includegraphics[width=0.8\textwidth]{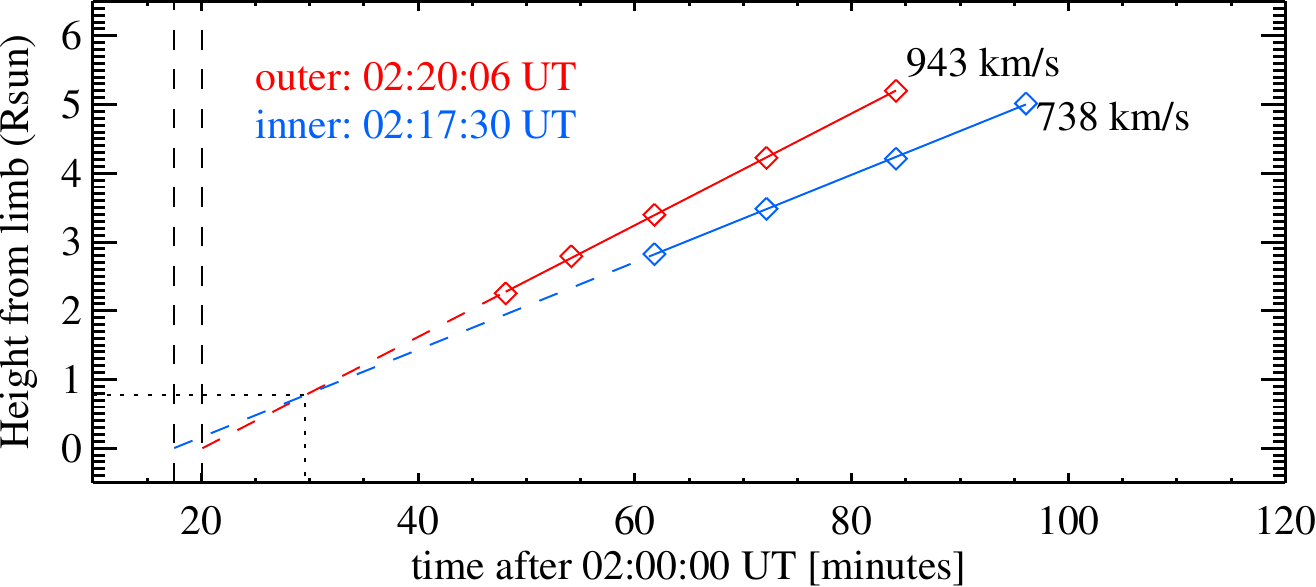}
    \caption{Top: LASCO/C2 base difference images (02:24:05 UT image is the base). White circle represents the solar limb and the white cross symbol marks the location of NOAA AR 12371. Red and blue diamonds mark the outer and the inner edge of the CME at each time instance.
    Bottom: CME height profiles measured along the trajectory of diamond symbols. The outer CME edge time profile is in red color and that of the inner CME edge is in blue. The apparent CME onset time at the limb is estimated to be 02:17:30~UT for inner CME and 02:20:06~UT for the outer CME. The interaction of these two CMEs is estimated to occur at the height of 0.77~R$_{\bigodot}$ at 02:29:36~UT.}\label{lasco_cme}
\end{center}
\end{figure}

A halo CME was detected by the LASCO/C2 coronagraph on June 21 at 02:36:05~UT. Figure~\ref{lasco_cme} shows a series of base difference images calculated using the 02:24:05~UT image as a base (not shown in the figure), with the location of the source AR marked by a cross. A classical three-part structure is distinctive in this halo CME: a bright front pointed by a red diamond, an inner bright core (blue diamond), and a dark cavity between them. From the 03:24:05~UT image, yet another core structure can be seen at the position angle of $\sim160\degr$. Such a double-core structure may be due to the projection effect. It is interesting, however, that the position angle of each core is consistent with the expansion direction of the coronal loops that we detected in the AIA~193~\AA\ data for Flare I. Also, recall that the main dimmings that appeared after Flare II were located north and south of the flaring AR. We calculated a height-time diagram for both the outer bright front and the inner core, and extrapolated it back to estimate the CME apparent onset time at the solar limb. If we assume that the CME bright front is associated with the dense coronal plasma shell and the core is associated with the prominence or flux rope material ejected with the CME \citep{Ril08, Wu99}, any onset time for the inner core earlier than onset time for the outer front appears contradictory at first. However, the x-t diagram in Figure~\ref{x-t diagram} shows that the second ejecta that started after Flare II is faster than the first ejecta, and is likely to overtake the first one in the low corona as it propagates. It can be suggested that in addition to the two-step flare and field opening, the ejection was a two-step process as well with the faster ejecta following in the wake of the first slower eruption. The bottom panel in Figure~\ref{lasco_cme} shows that the projected propagation speed of the inner core (\kms{738}) is lower than that of the outer bright front (\kms{943}), and thus its apparent onset time at the limb was a few minute earlier than that of the outer front. Even if we assume a linear CME propagation speed, the real onset time may be earlier than the estimate since the AR location was near the disk center (-220\arcsec, 158\arcsec). The earlier core onset may be due to the coronal loop expansion that was initiated in Flare I, while the later onset time of the bright front can be associated with the main plasma eruption during Flare II.

\begin{figure}[tb]
\begin{center}
    \includegraphics[width=0.95\textwidth]{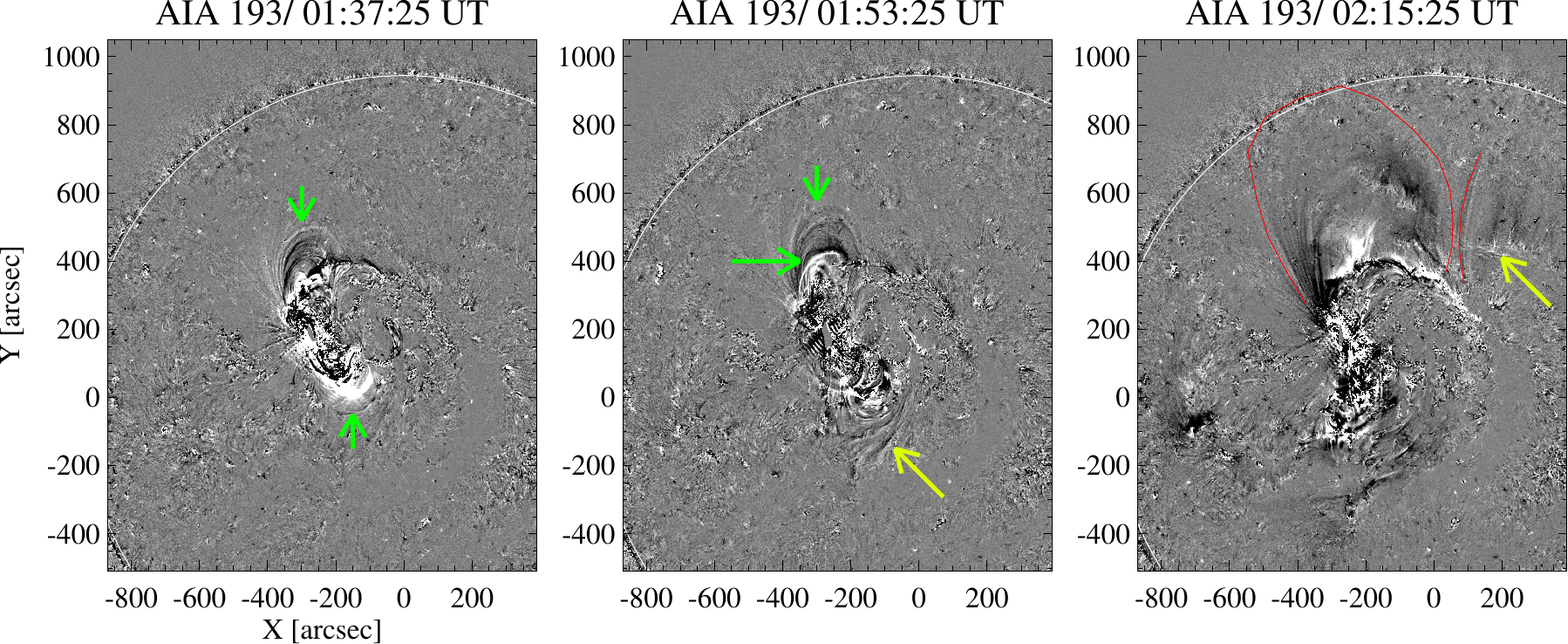}
    \includegraphics[width=0.3\textwidth]{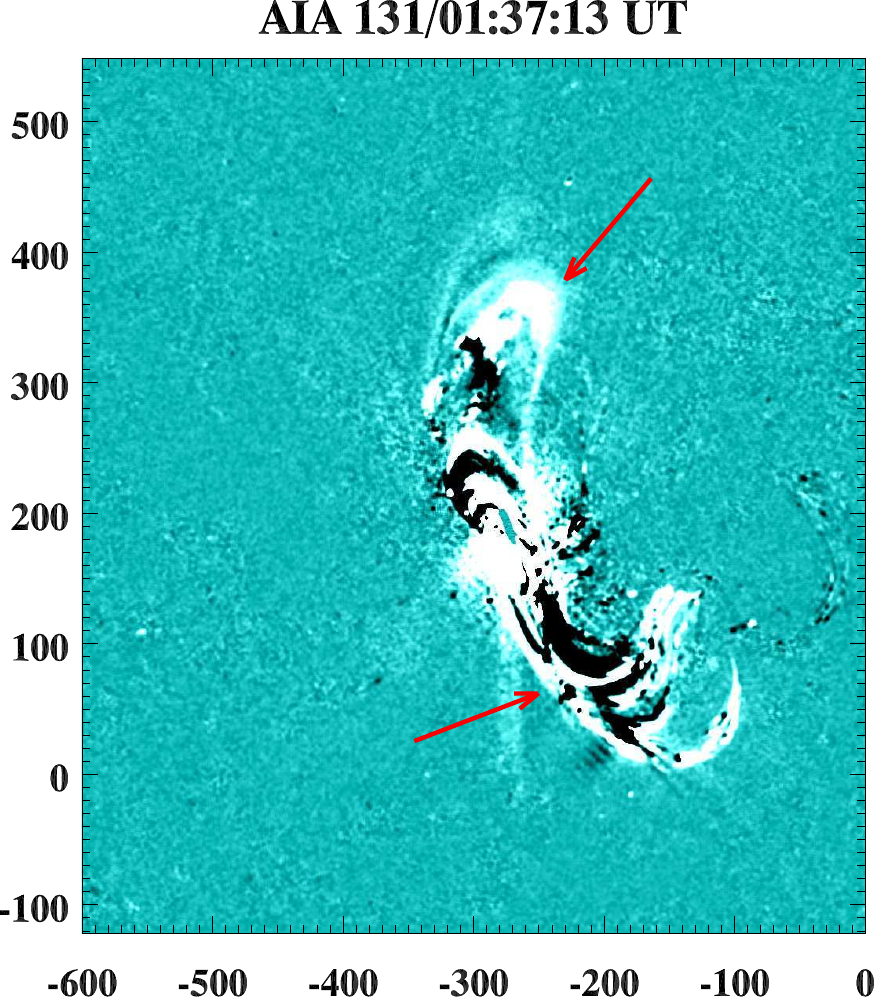}
    \includegraphics[width=0.3\textwidth]{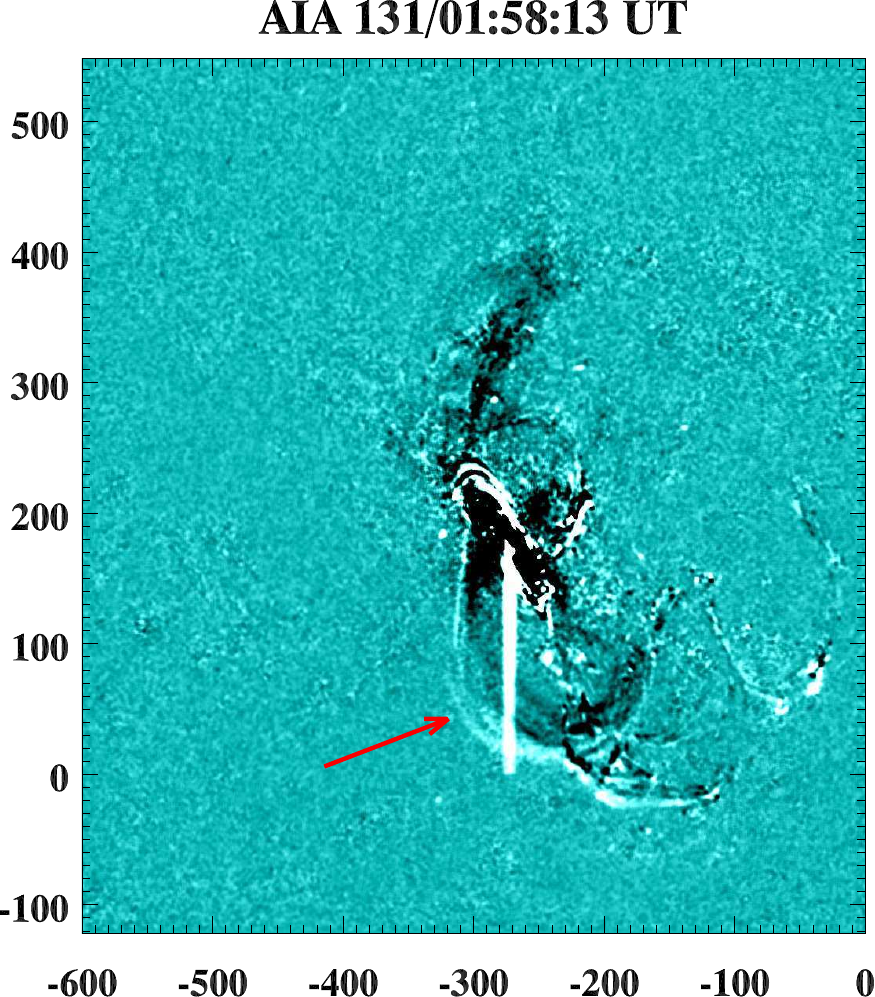}
    \includegraphics[width=0.3\textwidth]{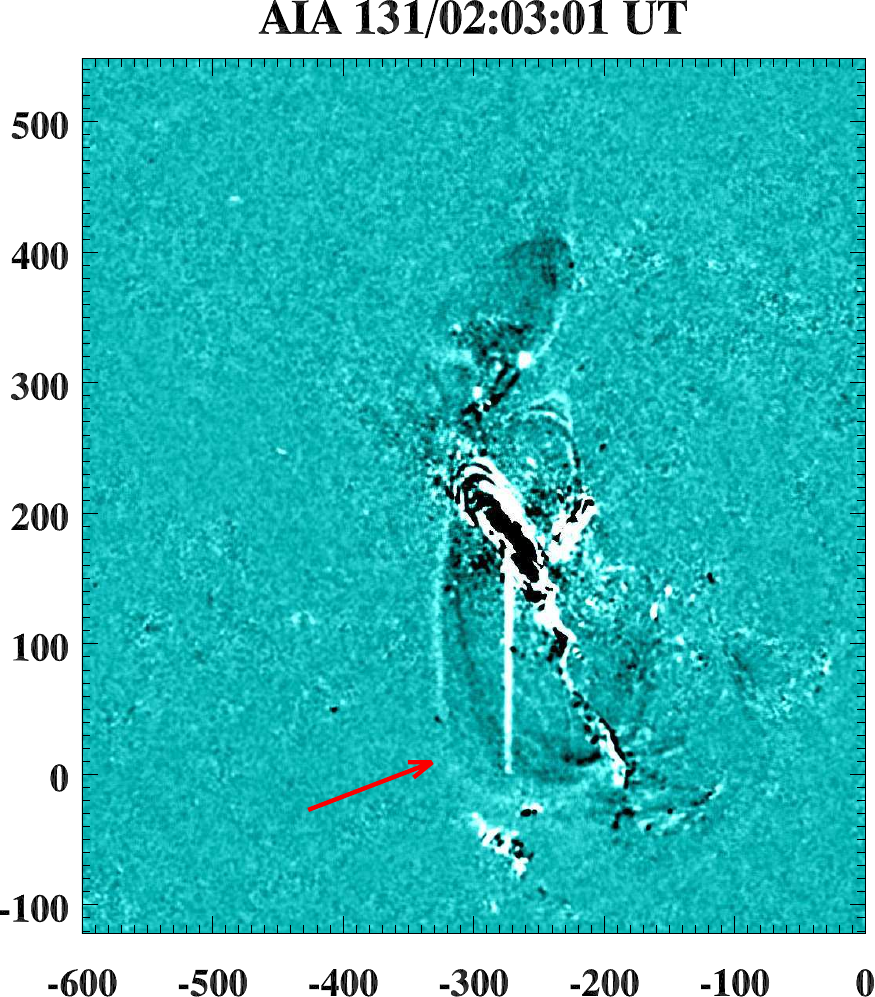}
    \caption{Running difference images of AIA~193~\AA\ (upper) and AIA~131~\AA\ (lower) during Flare I and II. Expanding coronal loops are pointed by green (red) arrows, and the disturbed neighboring loops in the southern region are pointed by a yellow arrow (upper). Expanded loops in the upper panel are outlined by red curves.}
\label{full_diff}
\end{center}
\end{figure}

Figure~\ref{full_diff} shows AIA~193~\AA\ and AIA~131~\AA\ running difference images where the loops of interest, pointed by short green arrows, are seen slowly expanding (left panels), but they did not fully erupt during Flare I (before 02:06:00~UT). Some loops show contraction (pointed by long green arrow at 01:53:25~UT, upper panel) indicating that this first eruption was not successful probably due to the overlying fan-shaped structure. Also in the southern part of the AR, the expanding loops experienced continuous interactions with the surrounding loops making them disturbed (middle panels). Neither field opening nor strong outflows were detected in the southern part at this time. On the other hand, the upper-right panel (02:15:25~UT) shows a partially circular bright-dark disturbance in the southern part just after Flare II. Although the signal is not that strong, such a structure is similar to the EUV wave indicative of a CME eruption. In the northern part the expanding loops accelerated and were followed by a significant plasma outflow. Moreover, a quiescent filament in the north-west of the flaring region at (200\arcsec, 410\arcsec) showed transverse oscillations a few minutes after Flare II (yellow arrow). At the same time, a lateral displacement of open field lines was also detected (short red curve). This sequence of events strongly suggest that the main eruption of coronal magnetic fields occurred in association with Flare II, which can only be possible only if the overlying fan-shaped field is opened.

\subsection{Radio bursts associated with flares}
To investigate signatures of particle acceleration during both flares, we analysed radio data obtained from Nobeyama Radioheliograph \citep[NoRH;][]{Nak94}, Nobeyama Radio Polarimeters \citep[NoRP;][]{Nak85}, and Learmonth solar observatory of Radio Solar Telescope Network (RSTN).

\begin{figure}[tb]
\begin{center}
    \includegraphics[width=9.6cm]{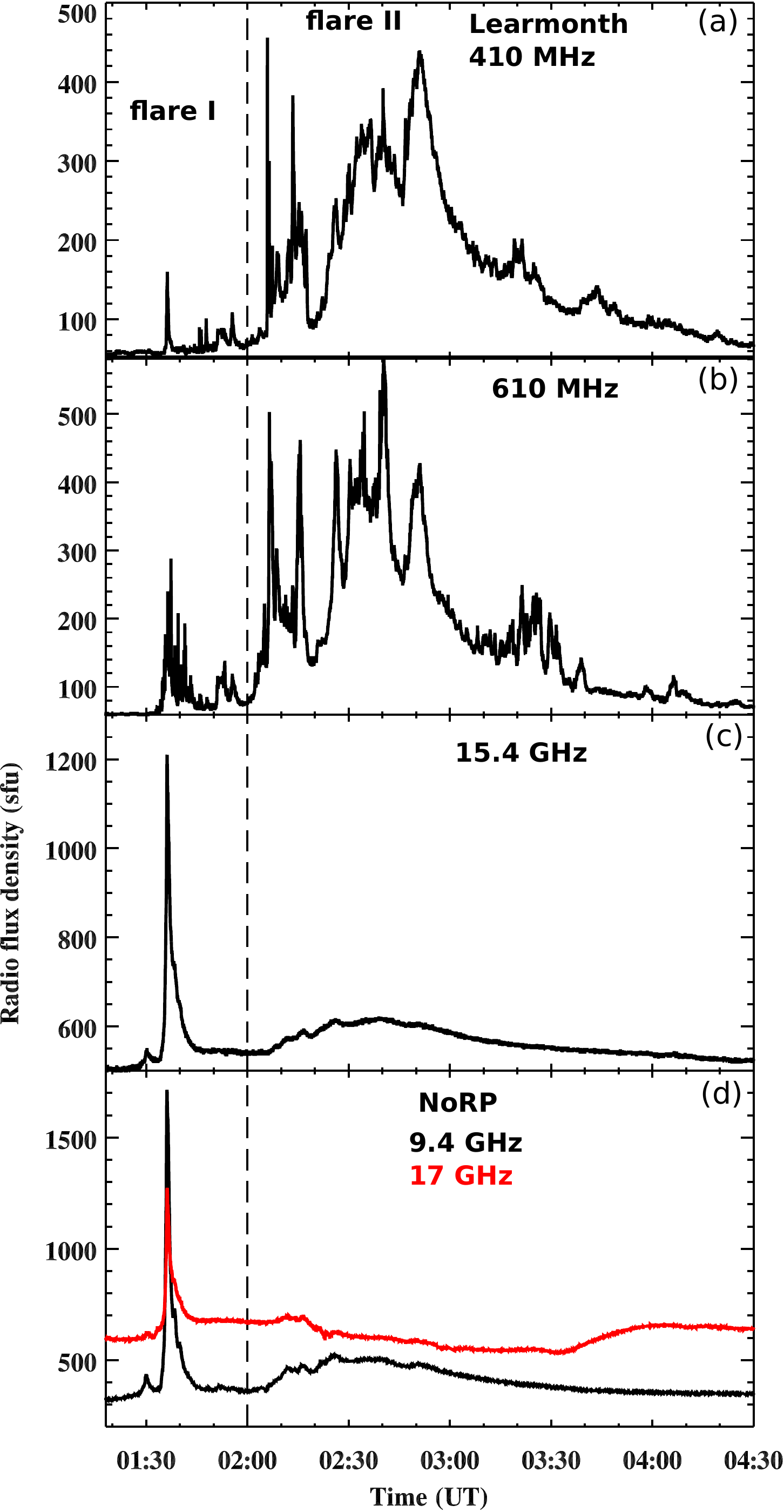}
    \includegraphics[width=6.2cm]{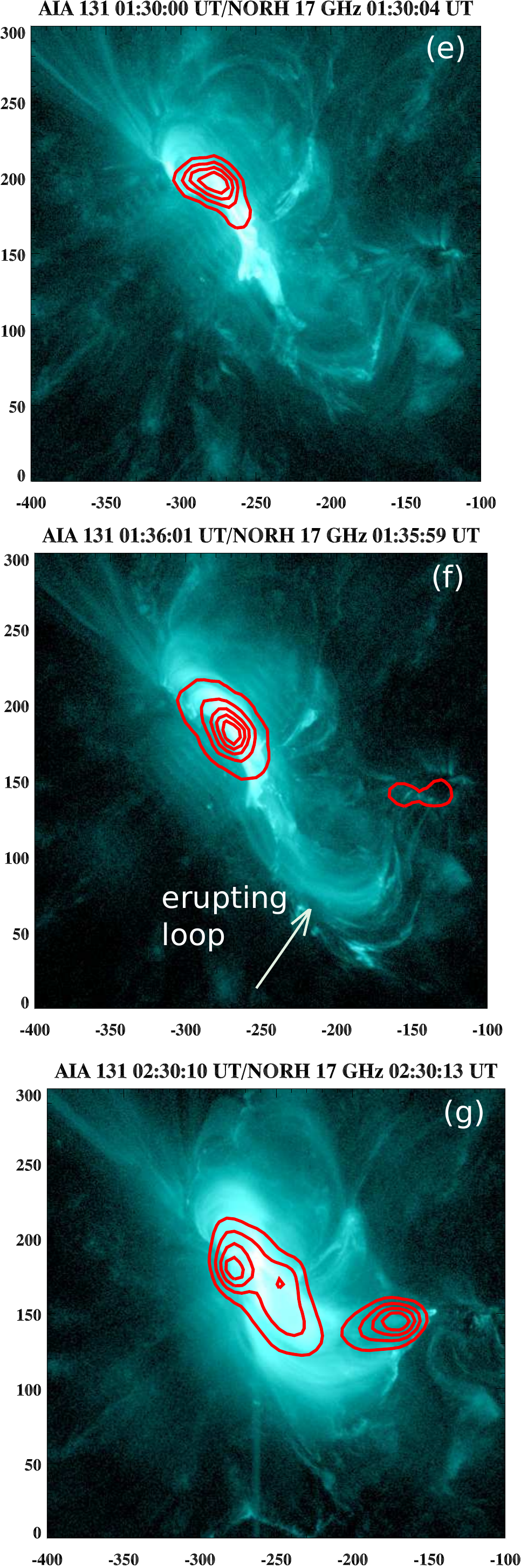}
    \caption{(a-c) Radio flux density (sfu) profiles in 410, 610, and 15400 MHz observed during both flares. (d) 9.4 and 17 GHz flux profiles from NoRP. (e-f) NoRH 17 GHz brightness temperature (T$_B$) contours overlaid on AIA 131 \AA~ images. The contour levels are 20$\%$, 40$\%$, 60$\%$, 80$\%$ of the peak T$_B$. X and Y axes are labeled in arcsecs.}\label{aia131}
\end{center}
\end{figure}

Figure~\ref{aia131}(a-c) displays 1 sec cadence radio flux density profile (in sfu) in 410, 610, and 15400~MHz during 01:18-04:30~UT observed at the Learmonth radio station. The decimetric emission is assumed to be tracing nonthermal electrons accelerated from the current-sheet during a magnetic reconnection process \citep[e.g.,][]{Ben11, Kum17a}. The emission mechanism is plasma emission, where the radio source height (in corona) depends on the plasma density. Interestingly, the decimetric bursts (410 and 610 MHz) reveal a weak emission during the Flare I (01:36 UT), but much stronger quasi-periodic emission (until 03:00 UT) during the Flare II. However, the weak decimetric emission persists until almost 04:00 UT.

The microwave emission is a type of the gyrosynchrotron emission \citep{Dul85}. Figure~\ref{aia131}(d) shows 9.4 and 17 GHz flux profiles from NoRP. During Flare I, we noticed an impulsive microwave burst (9.4, 15.4, 17 GHz) at $\sim$01:36 UT. However, only a weak gradual microwave emission was detected during the second flare (02:00-3:30 UT).

To determine the location of the microwave emission, we overlaid the NoRH 17 GHz brightness temperature (T$_B$) contours over the co-temporal AIA~131~\AA\, images (Figure~\ref{aia131}(e-g)). We observed AIA~131~\AA\, brightening of the sheared arcade in the core of the AR at 01:30 UT. And the small bump in the microwave flux profiles at 01:30 UT and the 17 GHz source seems to originate from the sheared arcade. The strong microwave emission at 01:36 UT (Figure~\ref{aia131}(f)) originated along with the appearance of the AIA~131~\AA\, bright loops (marked by arrow) from the same place along with a formation of a weak source (3$\%$ of the peak T$_B$) near the opposite footpoint of the loop. This source becoame more prominent at 02:30~UT during Flare II (Figure~\ref{aia131}(g)). The appearance of the southern hot loops (at 01:36 UT) indicates the ongoing magnetic reconnection in the core as well as trapping of energetic electrons in the strong core field region (i.e., confined flare).

The strong quasi-periodic decimetric emission during the second flare indicates repetitive acceleration of nonthermal electrons (outwards) in the corona and is a clear evidence of magnetic reconnection that persisted until $\sim$04:00 UT.

The RHESSI X-ray flux profiles also revealed the quasi-periodic emission in 6-12 keV (thermal) and 25-50 keV (nonthermal) emission during Flare I (i.e., 00:50-01:30 UT), and flare II (02:10-02:30 UT), which suggests that the repeated internal tether-cutting reconnection started at $\sim$00:50 UT during Flare I and continued until $\sim$01:36 UT. The internal tether-cutting reconnection among the sheared arcade possibly produced the erupting structure that was detected only in the AIA~131~\AA\, channel.
The hot flux ropes are generally produced by the reconnection of the sheared arcades and have been detected previously in the AIA~131~\AA\, channel \citep{Kum14, Yur15}.

\section{SUMMARY AND DISCUSSIONS}
Using the SDO/AIA~304~\AA\, data, we studied the formation of a strikingly large secondary flare ribbon associated with two successive flares that occurred within an one-hour time interval. We found the following: 1) the observed secondary flare ribbon was quasi-circular and it partly encircled the AR and the associated quiet Sun; 2) it spanned 500\arcsec\, in the east-west direction and 650\arcsec\, in the north-south direction; 3) the ribbon exhibited subsequent brightenings propagating in the clockwise direction with the estimated speed at about \kms{160} and unusual radial contraction in the northern part; and 4) radial ribbon expansion in its southern part proceeded at the rates of \kms{2.6} to \kms{13.6}.

In addition to the dynamical properties of the QCSR, various phenomena related to the eruption process were also well observed during the entire event, such as: 1) the distance between two ribbons abruptly increased from 60\arcsec\, to 140\arcsec\, at the Flare II onset; 2) the coronal loop expansion initiated with Flare I and accelerated after Flare II; 3) the main dimmings developed near both the north and south footpoints of the AR sigmoid; 4) the arc-shaped dimming appeared at the outer edge of the QCSR brightening; and 5) the CME inner core moving faster than the CME outer front.

Our observations indicate that the secondary ribbon is not very much different from a normal flare ribbon in the way that they both are the result of magnetic reconnection. The apparent differences in their shape and size can be attributed to the different underlying magnetic configurations. The magnetic field extrapolation revealed that the quasi-circular shape of the ribbon is due to the magnetic field topology that includes fan shaped elements, especially in the northern part of the ribbon. The subsequent brightening propagating in the clockwise direction may be explained in the same way as in the case of the circular ribbon with an asymmetric fan field \citep{Mas09}. However, the time interval between the onset of Flare II and the brightening at each ribbon location is too large to be explained in terms of the electron precipitations from a 3-D null point. Using the null point location calculated from the PFSS model, we estimated the travel speed by taking the ratio between the distance from the null point to the QCSR brightening and the elapsed time from Flare II. The estimated speed was about \kms{2400} in the northern part of the QCSR, and it is even lower in the southern part, which are one or two orders of magnitude lower than those required by the electron precipitation hypothesis. We therefore suggest that the clockwise subsequent brightening can be interpreted in terms of the slipping reconnection between the expanding sigmoid field and the overlying fan-shaped field. Note that the slipping reconnection between erupting and overlying structure will lead to quasi-periodic acceleration of nonthermal electrons, which is seen in the decimetric radio emission.

The two main dimmings formed in the northern and souther parts of the AR are also the evidence of reconfiguration of the magnetic field associated with the large-scale magnetic reconnection between the expanding structures and the overlying fields. A similar dimming associated with reconnection of a sigmoid with the overlying transequatorial loops has been previously reported by \citet{Man96}. However, they did not detect a QCSR or fan-shaped loops. In addition to the main dimmings, another arc-shaped dimming was detected just outside of the QCSR's bright edge which further supports the idea of the reconnection between erupting structures and overlying fields. \citet{Man07} suggested that additional dimmings may occur along the footpoints of an overlying field when a flux rope is erupting and interacting with this overlying field. Their detected dimming was located all around an AR rather than concentrated at the footpoints of a sigmoid. The arc-shaped dimming in the present work is consistent with their scenario.

While the northern ribbon brightening may be explained by the slipping reconnection under the partial fan-shaped field structure, a different mechanism seems to play a role in the southern part of the QCSR. The magnetic field connecting this part to the AR is no longer fan-shaped, but it is arcade-shaped as was confirmed from the field extrapolation. In addition, the ribbon expansion and the subsequent brightenings normal to the ribbon spine were not expected in the fan-spine reconnection model. The characteristics of the south part of the QCSR are more similar to those of parallel ribbons rather than circular ones. Although the exact mechanism for this part is unclear, it seems that the interaction between the expanding sigmoid ambient fields and the arcade field of the QCSR's south part may have played a role. We detected a clear eruption of a hot structure in the AIA~131\AA\, channel, along the south-east direction, associated with the formation of the QCSR. A secondary reconnection at QSLs driven by either waves \citep{Ram66, Zha14} or interaction of magnetic fields \citep{Tor11,Liu15, Sch11, Lee16} is often considered to be the mechanism for consecutive reconnections. The radial outward expansion of the QCSR brightening in combination with the apparent quick recovery may be a similar behavior of the typical separation of parallel ribbons.

The scale of the observed QCSR is the largest ever detected. Normally, circular ribbons have been observed in relatively small ARs with the parasitic polarity surrounded by opposite polarity fields. Then how such a large ribbon with a well-defined quasi-circular shape could be formed? The magnetic field distribution can provide an answer. In our case, a unipolar spot, N2, was located between the negative polarity, N1, of the eastern bipolar spots and the positive polarity of the western ambient field. Such distribution was favorable for a fan shaped magnetic configuration that was a part of large-scale pseudo streamer.

Morphological changes of the ribbon provide important information on the eruption process. Recent statistical study of flare ribbons by \citet{Tor17} demonstrated that compared to confined flares, the eruptive flares showed a larger ratio of the ribbon area to AR area. According to the authors, this ratio reflects the relative scale of the magnetic field involved in reconnection compared to the ambient field that is an obstacle for fast and powerful eruption. The sudden increase of the distance between parallel flare ribbons just after Flare II is thus consistent with their finding. The QCSR discussed here was not taken into account in their analysis even though the same AR was included in their survey. Taking into account this large-scale secondary ribbon will increase the resulting ratio even more thus reinforcing our suggestion that the CME was initiated during Flare II.

In addition, the main factor that determines the size of a flare ribbon is the scale of the magnetic field involved in flare reconnection. Our study on the QCSR that has developed between an AR and the background magnetic field suggests that the interaction between the AR and the ambient fields should be considered in the understanding of the eruption process. In majority of previously reported eruptions, not much attention was paid to the role of ambient fields, especially in case of single eruptions. On the other hand consecutive eruptions either in a single AR or in different regions, (commonly called sympathetic eruptions) reveal global connection of the magnetic field and give us an important clue to the eruption process \citep{Sch11}. The details of how one eruption can trigger another one at a significant distance are well demonstrated in the MHD simulations by \citet{Tor11}. By adopting a similar mechanism, it can be explained how the development of the observed QCSR during Flare II may have been initiated by Flare I sigmoid eruption.

\acknowledgments Authors are grateful to the anonymous referee for constructive comments that improved the manuscript. We thank M.L.~DeRosa for his help in analyzing PFSS model.
E.-K.L. and other KASI members are supported by the KASI ``Planetary system research for space exploration" grant. V.Y. acknowledges support from AFOSR FA9550-15-1-0322 and NSF AST-16114457 grants. P.K. is supported by an appointment to the NASA Postdoctoral Program at the Goddard Space Flight Center, administered by Universities Space Research Association through a contract with NASA. The work of the SNU team was supported by the Korea Astronomy and Space Science Institute under the R\&D program, Development of a Solar Coronagraph on International Space Station (Project No. 2017-1-851-00), supervised by the Ministry of Science, ICT, and Future Planning.
The SDO data were (partly) provided by the Korea Data Center (KDC) for SDO in cooperation with NASA, which is supported by the ``Development of Korea Space Weather Research Center" project of the Korea Astronomy and Space Science Institute (KASI). RHESSI is a NASA Small Explorer. Hinode is a Japanese mission developed and launched by ISAS/JAXA, with NAOJ as domestic partner and NASA and STFC (UK) as international partners. SOHO is a project of international cooperation between ESA and NASA. The Nobeyama RadioHeliograph and Polarimeters are operated by the International Consortium for the Continued Operation of Nobeyama. This work was partly carried out on the Solar Data Analysis System (SDAS) operated by the Astronomy Data Center in cooperation with the Solar Observatory of the National Astronomical Observatory of Japan.

\clearpage

\end{document}